\def\makeSM{1}
\documentclass[
prx,
twocolumn,
amsmath,
amssymb,
floatfix,
footinbib,
bibnotes,
longbibliography
]{revtex4-1}
\pdfoutput=1

\usepackage[colorlinks=true,citecolor=blue,linkcolor=magenta]{hyperref}
\usepackage{amsmath}
\usepackage{graphicx}
\usepackage{changepage}
\usepackage{fancyhdr}
\usepackage{amsthm, amssymb}
\usepackage{floatflt}
\usepackage{float}
\usepackage{array}
\usepackage[usenames,dvipsnames]{color}
\usepackage{mathtools}

\newcommand{\usenomenclature}{}
\ifdefined\usenomenclature
\usepackage[noprefix]{nomencl}
\fi

\newcommand{\beq}{\begin{equation}}
\newcommand{\eeq}{\end{equation}}
\newcommand{\eqn}[1] {eqn.~(\ref{#1})}

\newcommand{\fig}[1]{fig.~\ref{#1}}
\newcommand{\figSM}[1]{fig.~\ref{#1}}
\newcommand{\Fig}[1]{Fig.~\ref{#1}}

\newcommand{\ttd}[1]{#1}

\newcommand{\mylabel}[1]{\label{#1}} 
\newcommand{\mycite}[1]{\cite{#1}}

\newcommand{\bcen}{\begin{center}}
\newcommand{\ecen}{\end{center}}
\newcommand{\btab}{\begin{tabular}}
\newcommand{\etab}{\end{tabular}}
\newcommand{\bdes}{\begin{description}}
\newcommand{\edes}{\end{description}}
\newcommand{\mc}{\multicolumn}

\newcommand{\bea}{\begin{eqnarray}}
\newcommand{\eea}{\end{eqnarray}}

\newcommand{\bary}{\begin{array}}
\newcommand{\eary}{\end{array}}
\newcommand{\benum}{\begin{enumerate}}
\newcommand{\eenum}{\end{enumerate}}
\newcommand{\bitem}{\begin{itemize}}
\newcommand{\eitem}{\end{itemize}}

\newcommand{\ba} { \bm{a} }

\newcommand{\be} { \mbox{\boldmath $e$}}

\newcommand{\D}[1]{\mbox{d}{#1}}

\newcommand \SMcite[1]{\mycite{SM}, \ref{#1}}

\newcommand \seccite[1]{sec.~\ref{#1}}

\makeatletter

\newcommand{\Rmnum}[1]{\expandafter\@slowromancap\romannumeral #1@}
\makeatother

\newcommand{\Lc}{{N}}
\newcommand{\LPsi}{{M}}

\newcommand{\Gc}{{G_c}}
\newcommand{\GPsi}{{G_\psi}}
\newcommand{\Sigc}{{\Sigma_c}}
\newcommand{\SigPsi}{{\Sigma_\psi}}

\newcommand{\cH}{\mathscr{H}}

\newcommand{\ci}{\mathfrak{i}}

\newcommand{\dt}{\textup{d}t}
\newcommand{\dw}{\textup{d}\omega}

\newcommand{\cb}{\bar{c}}
\newcommand{\Psib}{\bar{\psi}}

\newcommand{\Tr}{\textup{Tr}}

\renewcommand{\Im}{\mathrm{Im}}

\newcommand{\dz}{\mathrm{d}z}
\newcommand{\tbd}[1]{}
\newcommand{\coT}{\mathcal{T}}
\newcommand{\rT}{\mathbf{t}}
\def\action{\mathcal{S}}
\newcommand{\concept}[1]{}
\newcommand{\figconcept}[1]{}
\newcommand{\eqnconcept}[1]{}
\newcommand{\tabconcept}[1]{}
\newcommand{\appconcept}[1]{}

\def\Zneq{\mathcal{Z}_\textup{neq}}
\def\ctimeint{\int_\mathcal{C}\textup{d}}

\def \showfigures{1}

\newcommand{\ee}{\end{align}}

\def \Tr{\mathrm{Tr}}

\def \be{\begin{equation}}
\def \ee{\end{equation}}
\def \ba{\begin{array}}
\def \ea{\end{array}}
\def \bea{\begin{eqnarray}}
\def \eea{\end{eqnarray}}

\def \cH{{\cal{H}}}

\def \a{{\alpha}}

\def \b{{\beta}}

\def \D{{\Delta}}

\def \ba{\begin{align*}}
\def \ea{\end{align*}}

\def \mr{\mathrm}

\def \mc{\mathcal}

\newcommand {\apgt} {\ {\raise-.5ex\hbox{$\buildrel>\over\sim$}}\ }
\newcommand {\aplt} {\ {\raise-.5ex\hbox{$\buildrel<\over\sim$}}\ }

\newcommand{\ttdn}[1]{#1}

\def \titlename {Quench, thermalization and residual entropy across a non-Fermi liquid to Fermi liquid transition}
\def \authornames{Arijit Haldar$^{1,2}$, Prosenjit Haldar$^{1,3}$, Surajit Bera$^1$, Ipsita Mandal$^{4,5}$, and Sumilan Banerjee$^1$}
\def \affiliations{$^1$Centre for Condensed Matter Theory, Department of Physics, Indian Institute 
of Science, Bangalore 560012, India\\
$^2$Department of Physics, University of Toronto, 60 St. George Street, Toronto, Ontario, M5S 1A7, Canada\\ 
$^3$Laboratoire de Physique Th\'{e}orique, IRSAMC, Universit\'{e} de Toulouse, CNRS, UPS, France\\
$^4$Laboratory of Atomic and Solid State Physics, Cornell University, Ithaca, NY 14853, USA\\
$^5$Faculty of Science and Technology, University of Stavanger, 4036 Stavanger, Norway
}

\begin{document}

\title{\titlename}
\author{\authornames}
\affiliation{\affiliations}

\date\today

\begin{abstract}
We study the thermalization, after sudden and slow quenches, in an interacting model having a quantum phase transition from a Sachdev-Ye-Kitaev (SYK) non-Fermi liquid (NFL) to a Fermi liquid (FL). The model has SYK fermions coupled to non-interacting lead fermions and can be realized in a graphene flake connected to external leads. A sudden quench to the NFL leads to rapid thermalization via collapse-revival oscillations of the quasiparticle residue of the lead fermions. In contrast, the quench to the FL shows multiple prethermal regimes and much slower thermalization. In the slow quench performed over a time $\tau$, we find that the excitation energy generated has a remarkable intermediate-$\tau$ non-analytic power-law dependence, $∼ \tau^{-\eta}$ with $\eta<1$, which seemingly masks the dynamical manifestation of the initial residual entropy of the SYK fermions. Our study gives an explicit demonstration of the intriguing contrasts between the out-of-equilibrium dynamics of a NFL and a FL in terms of their thermalization and approach to adiabaticity.

\end{abstract}

\maketitle

\section{Introduction}
One of the major frontiers in condensed matter physics is to describe gapless phases of interacting fermions without any quasiparticles, namely non Fermi liquids (NFL) \mycite{SachdevBook}. Recently, new insights about fundamental differences between NFLs and Fermi liquids (FL) have  been gained in terms of many-body quantum chaos and thermalization. This new impetus has come from exciting developments in a class of NFLs described by Sachdev-Ye-Kitaev (SYK) model, \cite{Sachdev1993,KitaevKITP,Maldacena2016} and its extensions \mycite{Gu2017,BanerjeeAltman2016,SJian2017b,Song2017,Davison2017,Patel2018,
Chowdhury2018,Arijit2017,Haldar2018}, and their connections with black holes in quantum gravity \mycite{Sachdev2010,KitaevKITP,Sachdev2015}. In particular, the model proposed in ref.\mycite{BanerjeeAltman2016} classifies the SYK NFL and a FL as two distinct chaotic fixed points, separated by a quantum phase transition (QPT). In this characterization, the NFL thermalizes much faster than the FL, as quantified by a rate of the onset of chaos or the Lyapunov exponent \mycite{Maldacena2015,KitaevKITP,BanerjeeAltman2016}.

However, the Lyapunov exponent is computed from an equilibrium dynamical correlation, the so-called out-of-time-ordered correlator \mycite{KitaevKITP,Maldacena2016,Kitaev2018}. Here, using the model of ref.\mycite{BanerjeeAltman2016} as a template, we ask whether such contrast between the NFL and FL persists even for thermalization from a completely out-of-equilibrium situation, e.g. a quantum quench. The model has two species of fermions, interacting SYK fermions coupled to another species of otherwise non-interacting fermions, referred to as \emph{lead} fermions. A QPT between a strongly interacting NFL and weakly interacting FL phases can be tuned in the model at low energies by varying the ratio, $p$, of numbers of sites on which the two types of fermions reside. Remarkably, the solvable nature of the model allows us to study its full non-equilibrium evolution after a quench exactly. By using non-equilibrium Keldysh field theory in the thermodynamic limit, as well as numerical exact diagonalization (ED) for finite systems, we demonstrate a drastic difference in thermalization rates for the NFL and FL after a sudden quench. In addition, we show that the quasiparticle residue of the lead fermions exhibits a dynamical transition as function of $p$ from \emph{collapse-and-revival oscillations} to \emph{prethermalized plateaus} as a function of time. This dynamical transition is similar to that seen in the interaction quench of Hubbard model \mycite{Eckstein2009}.

Furthermore, the Landau description of a FL is based on the concept of adiabatic time evolution from a non-interacting system under slow switching on of the interaction, without encountering a phase transition. 
Is it possible to evolve an NFL adiabatically to the FL and {\it vice versa}? We argue that such evolution is not possible here due to another intriguing aspect of the SYK NFL, namely the finite zero-temperature residual entropy (density) $S_0$ \cite{KitaevKITP,Sachdev2015,Maldacena2016}. The entropy is related to the Bekenstein-Hawking entropy of the black hole in the dual gravity theory \cite{KitaevKITP,Sachdev2015,Maldacena2016,Kitaev2018} and has relevance for strange metallic states described by local quantum criticality \cite{Gu2017,Davison2017,Chowdhury2018,Haldar2018}. We probe the signature of this entropy in the heat generated during non-equilibrium dynamics and characterize how the putative adiabatic limit is approached in the two phases, and across the QPT, after a slow quench with a finite rate. We show that, remarkably, the heat or energy of excitations $\Delta E$ \ttdn{generated} by the quench scales as $\Delta E(\tau)\sim \tau^{-\eta}$ with quench time $\tau$. Moreover, we find a direct manifestation of the equilibrium QPT in the scaling exponent $\eta$. We also contrast all the above results for the interacting model with those for an analogous non-interacting model under identical quench protocols. In particular, we show that the two models show drastically different thermalization behaviors in the thermodynamic limit.

The model studied here, could be realized in a graphene nano flake \cite{Can2019} attached to leads, and under a magnetic field. The  QPT also has close parallel in the NFL to FL transition in the multichannel Kondo model \cite{Parcollet1997}. Moreover, the study of dynamics after a quench in our model, where no quasiparticle description exists in one of the phases around the QPT, allows us to probe hitherto unexplored regime of many-body quantum dynamics. This is complementary to the previous studies of dynamics after quench across a QPT in integrable models \cite{Essler2016,DAlessio2016} or, weakly interacting systems with well defined quasiparticles \cite{Moeckel2008,Moeckel2009,Eckstein2009}. The scaling laws mentioned above can not be explained by the usual Kibble-Zurek scaling \cite{Dziarmaga2010,Polkovnikov2011}, unlike that in integrable or weakly-integrable models \cite{Essler2016,DAlessio2016,Moeckel2008,Moeckel2009,Eckstein2009}. Although, there have been a few studies on non-equilibrium dynamics of the original SYK model  \cite{Eberlein2017,Sonner2017,Kourkoulou2017,Bhattacharya2018}, none of them addressed the issue of a quench across a non-trivial QPT.

The remainder of this paper is organized as follows. In \seccite{sec:Model}, we introduce the interacting and non-interacting  models and describe the quench protocols. Section~\ref{sec:Results} discusses the results for the non-equilibrium evolutions after slow and sudden quenches in the large-$N$ limit obtained using non-equilibrium Schwinger-Keldysh method. Some results for finite-$N$ obtained via ED studies in a few limiting cases are also discussed in this section. In \seccite{sec:Conclusion} we conclude with the implications and significance of our results. The details of the non-equilibrium large-$N$ formulations, equilibrium spectral properties of the models and some additional results on the slow quench in the non-interacting model are included in Appendices \ref{sec:neq_action}, \ref{sec:gap_closing} and \ref{subsec:excnI}. The analysis of the slow quench using usual adiabatic perturbation theory and the breakdown of adiabaticity in our model are discussed in Appendix \ref{subsec:powlawbreak}. Additional details of the numerical calculations, ED and the results are given in the Supplementary Material (SM) \cite{SM}.

\begin{figure}[h]
\centering
\includegraphics[width=0.4\textwidth]{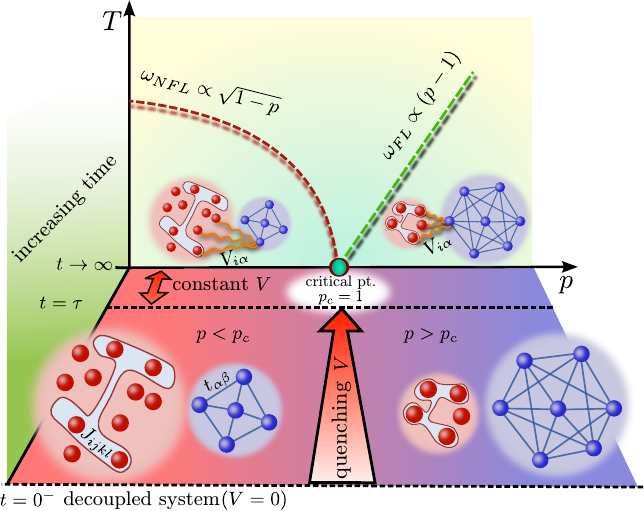}	
\caption{{\bf Model and the quench protocol:} SYK ($c$) fermions (red dots), interacting via random quartic coupling $J_{ijkl}$, on $N$ sites, are connected, at time $t=0$, to \emph{lead} ($\psi$) fermions (blue dots joined
by lines) using random-quadratic couplings $V_{i\alpha}$ with strength $V$. The $\psi$ fermions reside on $M$ sites and have random hopping amplitudes $t_{\alpha\beta}^\psi$. For a fixed site-fraction $p=M/N$,  the coupling $V$ is ramped from $0$ to a finite value over quench duration $\tau$, as depicted by a red arrow pointing into the page. For $t\to\infty$, the connected system is expected to relax to a thermal state in the equilibrium  $p-T$ phase-diagram, where $T$ is the final temperature. A critical point, $p_c=1$, separates the SYK NFL ($p<p_c$) and the FL ($p>p_c$). The low-energy NFL and FL behaviors persist up to the crossover scales $\omega_{NFL}$ and $\omega_{FL}$, respectively.}
\label{fig:Model}
\end{figure} 

\section{Model} \label{sec:Model}
\subsection{Interacting model}
As described schematically in fig.\ref{fig:Model}, we study a time ($t$)-dependent version of the model in ref.\cite{BanerjeeAltman2016}, $\mc{H}(t)=\mc{H}_c+\mc{H}_{\psi}+\mc{H}_{c\psi}(t)$, where
\begin{subequations} \label{eq.Model}
\bea
\mc{H}_c&=&\frac{1}{(2N)^{3/2}}\sum_{ijkl}J_{ijkl}c_{i}^{\dagger}c_{j}^{\dagger}c_{k}c_{l}
\label{eq.SYK}
\\
\mc{H}_{\psi}&=&\frac{1}{M^{1/2}}\sum_{\alpha\beta}t^\psi_{\alpha\beta}\psi_{\alpha}^{\dagger}\psi_{\beta}
\label{eq.Anderson}\\
\mc{H}_{c\psi}(t)&=&\frac{f(t)}{(NM)^{1/4}}\sum_{i\alpha}(V_{i\alpha}c_{i}^{\dagger}\psi_{\alpha}+V_{i\alpha}^{*}\psi_{\alpha}^{\dagger}c_{i}).
\label{eq.Coupling}
\eea 
\end{subequations}

The model (\fig{fig:Model}), has two species of fermions -- (1) the SYK fermions ($c$), on sites $i=1,\dots,N$, interacting with random four-fermion coupling $J_{ijkl}$ (\eqn{eq.SYK}), drawn from a Gaussian distribution with zero mean and variance $\overline{|J_{ijkl}|^2}=J^2$; and (2) the \emph{lead} fermions ($\psi$), on a separate set of sites $\alpha=1,\dots,M$ connected via random all-to-all hopping $t^\psi_{\a\b}$ (\eqn{eq.Anderson}). The SYK and the lead fermions  are quadratically coupled via $V_{i\a}$; $t^\psi_{\a\b}$ and $V_{i\a}$ are complex Gaussian random variables with zero mean, and variances $\overline{|t^\psi_{\a\b}|^2}=t_\psi^2$ and $\overline{|V_{i\a}|^2}=V^2$, respectively. 

The model is exactly solvable for $N,M\to\infty$ with a fixed ratio $p=M/N$, that is varied to go through the QPT between NFL and FL at a critical value $p=p_c=1$ \cite{BanerjeeAltman2016}. Two crossover scales, \ttd{$\omega_{NFL}$ and $\omega_{FL}$,} approach zero from either sides of the QPT (\fig{fig:Model}). The residual entropy density $S_0(p)$ of the SYK NFL continuously vanishes at the transition \cite{BanerjeeAltman2016}. This is one of the unique features of the QPT. 

To probe the non-equilibrium dynamics, we make the coupling term in \eqn{eq.Coupling} time dependent. In particular, we perform \emph{geometric} quenches (fig.\ref{fig:Model}) by switching on the coupling between the two initially disconnected subsystems (a) suddenly such that $f(t)=\Theta(t)$, the Heaviside step function, and (b) by slowly ramping up the coupling over a time $\tau$, i.e. $f(t)= r(t/\tau)[\Theta(t)-\Theta(t-\tau)]$; $r(x)$ is a ramp function, e.g. $r(x)=x$. Before the quench, the disconnected subsystems, with a preset \ttdn{site-ratio} $p$, are at their own thermal equilibria at initial temperatures $T_i^c$ and $T_i^\psi$. We take $T_i^c,~T_i^\psi\to 0$ so that SYK and lead fermions belong to the NFL and (non-interacting) FL states, respectively. As shown in \fig{fig:Model}, for $t\to \infty$, depending on whether $p<1$ or $p>1$, the coupled system eventually is expected to thermalize to either the NFL or the FL state, respectively. In any case, one of the subsystems always undergoes a transition, either from FL to NFL or {\it vice versa}, under the quench. 

\subsection{Non-interacting model}
To contrast the behavior of the above interacting model, and to demonstrate the crucial role of interaction in the non-equilibrium dynamics after quench and eventual thermalization, we also consider an analogous non-interacting model. The latter is obtained by replacing the interaction term for the $c$ fermions with a random hopping term similar to the one appearing in the $\psi$-fermion Hamiltonian [\eqn{eq.Model}]. To this end, we have $\mc{H}^{(NI)}(t)=\mc{H}_c+\mc{H}_{\psi}+\mc{H}_{c\psi}(t)$,
with 
\bea
\mc{H}^{(NI)}_c&=&\frac{1}{(N)^{1/2}}\sum_{ij}t^c_{ij}c_{i}^{\dagger}c_{j}.
\mylabel{eq.SYK_SI}
\eea
Here $t^c_{ij}$ is a complex Gaussian random variable with zero mean, and variance $\overline{|t^c_{ij}|^2}=t_c^2$; $\mc{H}_\psi$ and $\mc{H}_{c\psi}(t)$ are same as in \eqn{eq.Model}.

Below, we first briefly discuss the method for studying the time evolution of the systems, followed by the results for sudden and slow quenches.

\section{Results} \label{sec:Results}
{\bf Non-equilibrium evolution:} We use standard Schwinger-Keldysh non-equilibrium Green's function technique \cite{KamenevBook,StefanucciBook} to study the quenches described above. Utilizing the closed-time-contour Schwinger-Keldysh action (see Appendix \ref{sec:neq_action}) for the model we derive the  Kadanoff-Baym (KB) equations for the disorder-averaged non-equilibrium Green's functions,  $G_s^<(t_1,t_2)$, $G_c^>(t_1,t_2)$ ($s=c,\psi$),  e.g. $G_c^>(t_1,t_2)=-i\overline{\langle c_i(t_1)c_i^\dagger (t_2)\rangle}$; the overline denotes disorder averaging.
The KB equations are numerically integrated using a predicator-corrector scheme (see \SMcite{sec:pred_corr_KB}) starting from the initial equilibrium Green's functions 
(see \SMcite{sec:eq_Gfs}) for the disconnected system. The time-dependence of $\mc{H}(t)$ is encoded in KB equations via the local self-energies $\Sigma_s$, which could be exactly calculated in the large-$N$ limit for both the interacting and the non-interacting models (see Appendix \ref{sec:neq_action}). 

We obtain the time-dependent expectation value of an observable $\mc{O}(t)$, i.e. $\langle \mc{O}(t)\rangle\equiv \mr{Tr}[\rho(t) \mc{O}(t)]$ \ttdn{(see \SMcite{sec:neq_observable})}, using the Green's functions. Here $\rho(t)$ is the time-dependent density matrix and $\mc{O}(t)$ includes the explicit time dependence, if any, of the observables. To understand thermalization, we track how the contributions of the individual terms in \eqn{eq.Model} to the total energy $E(t)=\langle \mc{H}(t)\rangle$, e.g. $E_{c\psi}(t)=\langle \mc{H}_{c\psi}(t)\rangle$, relax after the quench. Since the whole system is isolated, we estimate the expected temperature $T_f$ of the putative thermal state at long times from the total energy $E(t)=E_f$, which is conserved after the quench.
\begin{figure*}
\includegraphics[width=\textwidth]{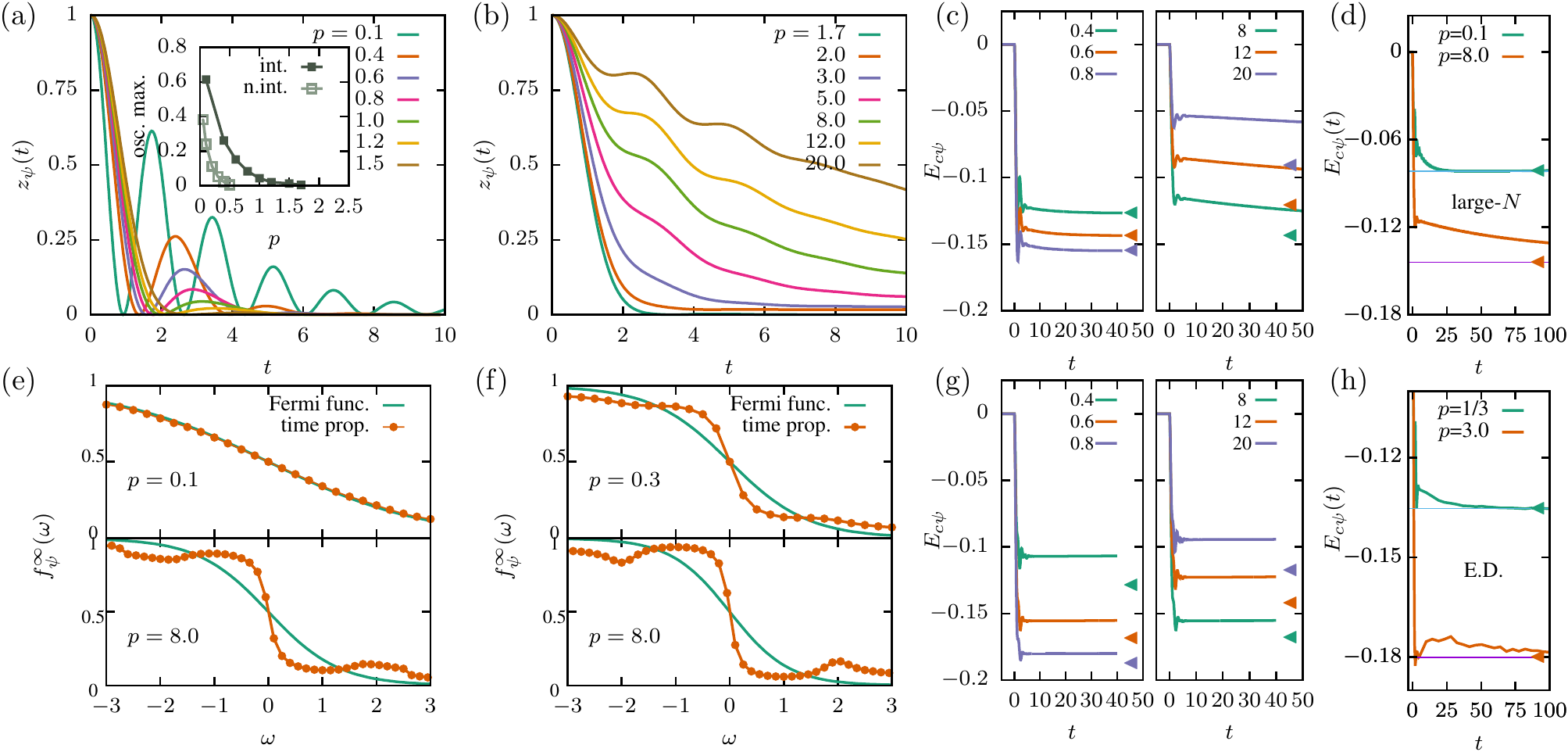}	
\caption{{\bf Dynamical transition and thermalization after a sudden quench:} 
(a) and (b) show the quasiparticle residue $z_\psi(t)$  for the ranges $p=0.1-1.5$ and $p=1.7-20.0$, respectively. In (a) inset, the curve labeled int. shows the height $z_1$ of the first maximum of the oscillations as a function of $p$. This is compared with the $p$-dependence of the oscillation maximum (marked as n.int.), when the $c$-fermions are made non-interacting. The effect of interactions pushes the critical value of  $p_c^{dyn}$ from $0.5$ to $1.5$.
Fig (c) gives the time evolution of bond energy $E_{c\psi}$, in the interacting model, for $p=0.4-0.8$ (left) and $p=8-20$ (right), while fig. (g) shows the time evolution in the non-interacting model for the same $p$ values. The $E_{c\psi}$ attain their respective equilibrium values (triangles), rather quickly, when $p<p_c^{dyn}$ in the interacting model and shows a slow approach to thermalization when $p\gg p_c^{dyn}$. On the other hand thermalization is completely absent, in the non-interacting model, and $E_{c\psi}$ never reach their respective equilibrium values regardless of the value of $p$.       
Sub fig.(e) shows the $\psi$-fermion long-time steady-state occupation function $f^\infty_{\psi}(\omega)$ (dots) in the interacting model, for $p=0.1$ (top), $8.0$ (bottom) respectively, and compares them with the Fermi function $n_\mr{F}(\omega,T_f)$ (line). Thermalization occurs for $p=0.1$, i.e $f^\infty_{\psi}(\omega)\to n_\mr{F}(\omega,T_f)$, while $p=8.0$ remains athermal. Fig.(f) shows the failure of $f^\infty_{\psi}(\omega)$(dots), for the non-interacting model, to approach $n_\mr{F}(\omega,T_f)$ (line) for both $p=0.3, 8.0$. Figs. (d), (h) give $E_{c\psi}$ as a function of $t$ obtained via large-$N$ calculation for $p=0.1,\ 8.0$ and via exact diagonalization (ED) for $p=1/3,\ 3.0$, respectively for the interacting model. The thermal expectation (diagonal ensemble) values for the large-$N$ (E.D.) case are shown by the arrow heads and horizontal lines.
}
\label{fig:SuddenQuench}
\end{figure*}

\subsection{Sudden quench} 
We first ask whether the contrast between dynamics of the NFL and FL can be seen even when the system is subjected to an abrupt non-equilibrium process. To address this, we study the case when the sub-systems are suddenly connected at $t=0$.
 We take $J=1$, $t_\psi=1$, $V=1$ and low initial temperatures, $T_i^c=0.05$ and $T_i^\psi=0$ \footnote{We keep the initial temperature of the SYK subsystem low but finite. Because of the divergent spectral density at $T=0$, the Green's function for SYK fermions in the initial equilibrium state can not be obtained numerically at strictly zero temperature.}. 
 The sudden quench leads to a rather high final temperature $T_f\sim 1$
  (see \SMcite{subsec:E_consv}).
 Before the quench, the lead fermions are non-interacting and the single-fermionic excitations are sharply defined at $T_i^\psi=0$. To track the quasiparticle evolution, we  compute an energy resolved time-dependent occupation, $n_\psi(\epsilon,t)=-iG_\psi^<(\epsilon;t,t)$, for the lead fermions. Here $\epsilon$ are the eigenvalues of the quadratic Hamiltonian in \eqn{eq.Anderson}, i.e. $\mc{H}_\psi=\sum_\epsilon \epsilon \psi_\epsilon^\dagger \psi_\epsilon$ and the Green's function $G_\psi^<(\epsilon;t_1,t_2)$ is obtained by integrating an appropriate KB equation (see \eqn{eq:sigma_from_G_contour_SYKq}).
 The quasiparticle residue, $z_\psi(t)=n_\psi(0^-,t)-n_\psi(0^+,t)$, is obtained from the occupation discontinuity at $\epsilon=0$. The vanishing of the residue indicates the destruction of the quasiparticles.

{\bf Collapse-and-revival oscillations and prethermal plateaus}: In our model, $z_\psi^\infty=z_\psi(t\to\infty)$ is expected to vanish for quench to any $p$ since the coupled system either thermalize to NFL or to a finite temperature FL state .
As shown in figs.~\ref{fig:SuddenQuench}(a)-(b), the collapse of the residue happens through two very different routes. First, for $p<p_c^{dyn}$, a critical value of $p$ corresponding to a dynamical transition, $z_\psi(t)$ undergoes collapse-and-revival oscillations \ttdn{(\fig{fig:SuddenQuench}(a))}
 Second, for $p>p_c^{dyn}$, $z_\psi(t)$ shows multiple long prethermal plateaus \ttdn{(\fig{fig:SuddenQuench}(b))}. For $V=1$, we find $p_c^{dyn}\approx 1.5$ from $z_1(p)\to 0$, where $z_1$ is the residue at the first maximum of the oscillations (see curve labeled int. in \fig{fig:SuddenQuench}(a) inset). Hence, the critical value $p_c^{dyn}$ for this dynamical transition is greater than the `equilibrium' critical ratio $p_c=1$. 
It is encouraging to find that similar oscillations and evidence of a dynamical transition have been observed \cite{Eckstein2009} in the interaction quench across Mott transition in the Hubbard model as well.

 However, in contrast to the interaction quench in Hubbard model, where the collapse-and-revival oscillations originate from on-site Hubbard repulsion, in our model, the oscillations are linked to a {\it soft} hybridization gap in the lead fermions. The gap appears due to hybridization (\eqn{eq.Coupling}) between the SYK and lead fermions, and closes at the NFL-FL transition (Appendix \ref{sec:gap_closing}). In fact, we observe a similar dynamical transition in the non-interacting model of \eqn{eq.SYK_SI} under an analogous sudden quench (see curve labeled n.int. in \fig{fig:SuddenQuench}(a) inset). The critical point in this scenario occurs at  $p_c^{dyn}\approx 0.5$. 
 We emphasize here that although a dynamical transition exists for the non-interacting case, the presence of interactions is crucial for thermalization (discussed in next section), which the non-interacting model fails to achieve for any value of $p$ (see \SMcite{sec:nI_sudden_results}). The SYK-interactions are also responsible for non-trivially shifting the critical value of $p_c^{dyn}$ from $0.5$ to $1.5$ as shown in \fig{fig:SuddenQuench}(a) (inset).
 
 {\bf Thermalization and long-time steady states:} The crucial aspects that distinguish the non-interacting 
 (\eqn{eq.SYK_SI}) and interacting (\eqn{eq.Model}) models, as well as the NFL and FL,  are the thermalization process and the long-time steady states. As shown in \fig{fig:SuddenQuench}(c)
  (also see \SMcite{sec:int_sudden_results},
   \figSM{fig:thermalization_int}(a)-(b)), $E_{c\psi}(t)$ reaches the thermal expectation corresponding to the temperature $T_f$ very rapidly for $p<p_c^{dyn}$, whereas there is a drastically slow, albeit finite, relaxation rate for $E_{c\psi}(t)$ towards the thermal value for $p>p_c^{dyn}$. 
 In contrast, $E_{c\psi}(t)$ does not relax to the expected thermal value for any $p$ in the non-interacting model, see \fig{fig:SuddenQuench}(g)
  (also \SMcite{sec:int_sudden_results},
   \figSM{fig:thermalization_nI}(a)-(b)). We further analyze the steady state through the Green's functions $G_s(\mc{T},\omega)=\int_{-\infty}^\infty G_s(t_1=\mc{T}+t/2,t_2=\mc{T}-t/2)e^{i\omega t}$, where $\mc{T}=(t_1+t_2)/2$. In the steady state, $G_s(\mc{T},\omega)$ becomes independent of $\mc{T}$. Moreover, for a thermal steady state at $T_f$, the steady-state occupation function $f_s^\infty(\omega)=\lim_{\mc{T}\to\infty}\ci G_s^<(\mc{T},\omega)/(2\mr{Im}G_s^R(\mc{T},\omega))$ should be equal to the Fermi function $n_\mr{F}(\omega,T_f)=1/(e^{\omega/T_f}+1)$, i.e. should satisfy the fluctuation-dissipation theorem (FDT). We find that for the interacting model (\eqn{eq.Model}), $f_s^\infty(\omega)=n_\mr{F}(\omega,T_f)$ (see \fig{fig:SuddenQuench}(e) top) for the NFL ($p<p_c^{dyn}$), whereas FDT gets violated (\fig{fig:SuddenQuench}(e) bottom) in the FL regime ($p>p_c^{dyn}$) for the largest $\mc{T}$ ($\sim 100J^{-1}$, where $J$ is the interaction strength in \eqn{eq.SYK}) accessed. 
 
We find that the FDT is never satisfied for the non-interacting model at any $p$, as seen in \fig{fig:SuddenQuench}(f) (also see 
\SMcite{sec:int_sudden_results},
  \figSM{fig:thermalization_nI}(c)-(d)). This is expected for the non-interacting case, where the long-time steady state is described by a generalized Gibbs ensemble (GGE) instead of the usual thermal Gibbs ensemble \ttd{\mycite{Polkovnikov2011,Essler2016,DAlessio2016,SM}}. Nevertheless, even in the interacting model, the FL phase shows an approximate GGE behavior by attaining a prethermal steady state within the time accessible in our numerical calculations. The prethermal GGE will presumably relax to a thermal state over a much longer time scale \cite{DAlessio2016,Stark2013}. Similar behaviors have been seen for quenches to FL phases in other interacting models \cite{Manmana2007}. In contrast, the strong interaction leads to rapid thermalization for the NFL phase. Hence the sudden quench in the interacting model demonstrates drastically different thermalization dynamics between the NFL and the FL phases.
  
 
  It is worthwhile to ask whether the contrast in the thermalization behaviors of NFL and FL persists even at finite $N$. In \fig{fig:SuddenQuench}(h), we show 
  	(see also \SMcite{sec:ED_sudden_results}) the results for $E_{c\psi}(t)$ obtained from ED studies of the model of \eqn{eq.Model} for $N=16$. The ED gives results similar to that at large-$N$, shown in \fig{fig:SuddenQuench}(g). Another pertinent question is whether the thermalization times of $E_{c\psi}(t)$ in NFL and FL phase can be directly related to their respective Lyapunov time scales ($\tau_\mr{L}$) \cite{BanerjeeAltman2016}. Here it is important to note that the sudden quench in our model leads to (see 
  	\SMcite{subsec:E_consv}, 
  	\figSM{fig:Tf_vs_p}) substantially high temperature $T_f\gtrsim J$. As a result, the relaxation of various high-energy modes also influence $E_{c\psi}(t)$, making  it hard to isolate $\tau_\mr{L}$ from the relaxation of high-energy modes. 
  	
We would also like to note that even though some specific low-energy features of SYK NFL and FL phases, like the temperature dependence of the Lyapunov time $\tau_\mr{L}$, cannot be 
ascertained from the dynamics after sudden quench, our results show that the low-energy fixed points for the initial states still drastically influence the thermalization process. This is despite the fact that the sudden quench leads to substantially high final temperatures.
  	

\begin{figure}[htb!]
\centering
\includegraphics[width=0.5\textwidth]{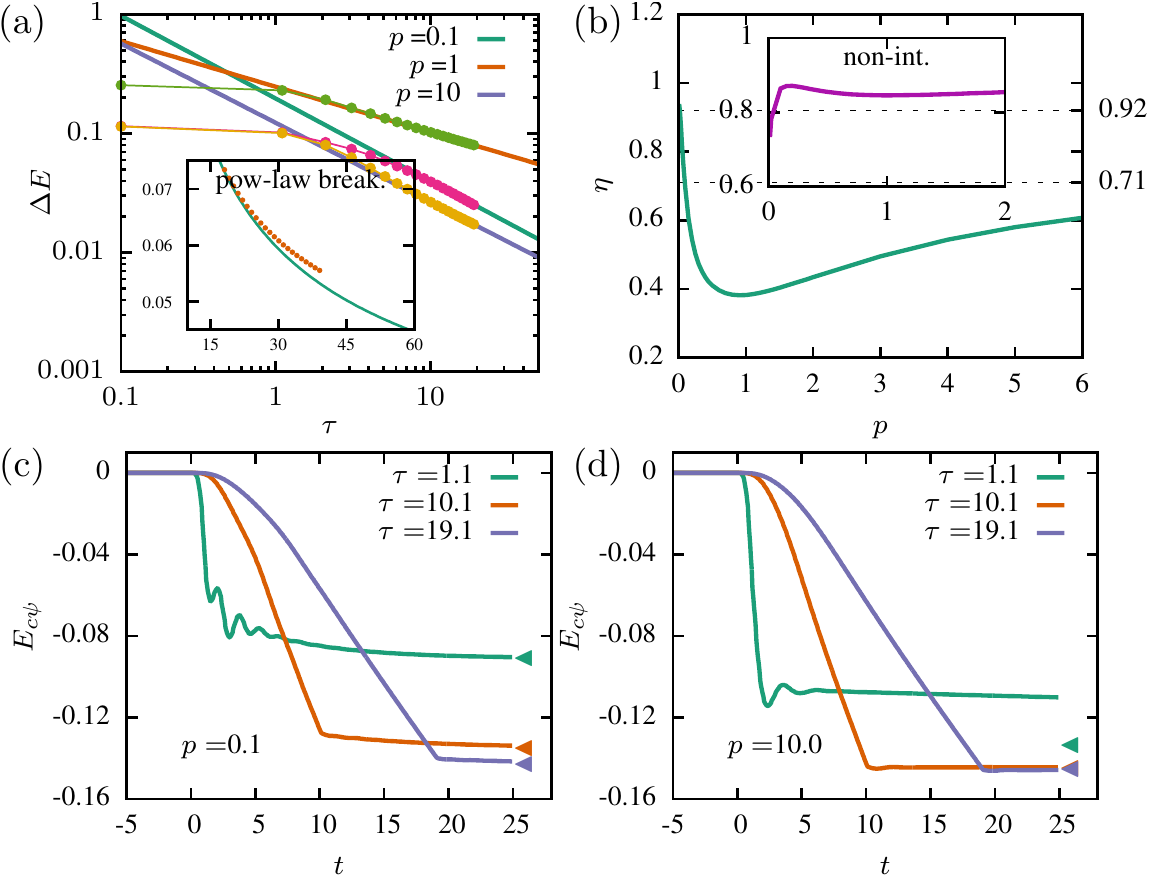}	
\caption{{\bf Heat production and thermalization in a slow quench:} (a) Heat or excitation energy $\Delta E(\tau)$ produced during the quench, as a function of quench duration $\tau$ (dots) and the power-law scaling fits, $\Delta E\sim \tau^{-\eta}$ (lines) for $p=0.1,1,10$. Inset shows an indication of the deviation from the power law for larger $\tau$ at $p=10$. (b) The exponent $\eta$ vs. $p$. The inset shows the $\eta(p)$ obtained for the quench in the non-interacting model \cite{SM}. (c) and (d) show time evolutions of $E_{c\psi}$ for $\tau=1.1,10.1,19.1$, at $p=0.1$ and $p=10$, respectively. The thermal expectations are shown by the arrow heads.}
\label{fig:SlowQuench}
\end{figure}

\subsection{Slow quench} 
We next address the question whether the initial decoupled NFL and FL subsystems can be adibatically evolved to the final states of the coupled system. To this end, we consider the slow quench where the coupling is changed slowly through a ramp, $r(t/\tau)$. Here we keep both the subsystems at some low initial temperature $T_i^c=T_i^\psi=T_i$ for $t<0$ and define the \emph{heat} or excitation energy \cite{Polchinski2016,Eckstein2010}, $\Delta E(\tau)=E(\tau)-\langle \mc{H}(\tau)\rangle_{T_i}$, produced during the quench; $\langle \mc{H}(\tau)\rangle_{T_i}$ is the thermal expectation of the final Hamiltonian $\mc{H}(\tau)$ at the initial temperature $T_i$. As shown in \fig{fig:SlowQuench}(a), remarkably, we find that $\Delta E(\tau)\sim \tau^{-\eta(p)}$ with $\eta(p)<1$, i.e.  a non-analytic power-law scaling. The exponent $\eta$ has a strong non-monotonic dependence on $p$ with a minimum around the QPT (\fig{fig:SlowQuench}(b)), revealing signatures of equilibrium QPT in the non-equilibrium evolution. We also find non-analytic scaling for the quench in the non-interacting model 
\ttdn{(see Appendix \ref{subsec:excnI})}. However, the exponent has a very weak dependence on $p$ (\fig{fig:SlowQuench}(b) inset). The particular non-analytic power laws cannot be explained through a standard adiabatic perturbation theory \cite{Polkovnikov2005,Dziarmaga2010,DeGrandi2010,Eckstein2010},
 as we show in Appendix \ref{subsec:powlawbreak}. Also, a Kibble-Zurek-type argument \cite{Dziarmaga2010,Polkovnikov2011} cannot be given for such a zero-dimensional system. We find the exponent to depend on ramp shape as well
 (see \SMcite{subsec:rampshapeseffect}). This is also not expected from adiabatic perturbation theory for an exponent $\eta<1$ \mycite{Eckstein2010,SM}. 

One possible promising route to understand the non-standard exponent $\eta(p)$ in the intermediate time window after the quench could be to construct a low-energy theory for the model of eqn.\eqref{eq.Model} along the line of Schwarzian theory for the pure SYK model \mycite{Maldacena2015,Kitaev2018}. A recent study \mycite{Almheiri2019} analyzes the quench dynamics using the Schwarzian theory for a SYK model suddenly coupled to a large thermal bath made out of another SYK model. However, the situation is somewhat more involved in our model due the strong back action of SYK fermions on the lead fermions in the NFL phase ($p<1$) and that of the lead fermions on the SYK fermions in the FL phase ($p>1$). Such back action is absent in the model of ref.\onlinecite{Almheiri2019} since the bath is infinitely larger than the system, and since both the bath and the system are described by SYK model. In our case, one needs to start with two uncoupled low-energy theories, one corresponding to the Schwarzian action for the SYK fermions with scaling dimension $\Delta=1/4$, and the other for the non-interacting fermions with scaling dimension $\Delta=1/2$. Hence the resulting theory after the quench, will not be that of the standard Schwarzian mode, but a different and more complicated one that includes the strong back actions among the subsystems. We would discuss this effective theory elsewhere \cite{Banerjee2020}.

As alluded earlier, for the quench to any finite $p$, e.g. from the NFL to FL, the residual entropy $S_0$ of the NFL, implies a violation of adiabaticity even for an arbitrary slow quench. The excitation energy $\Delta E(\tau)$ characterizes how $S_0$ metamorphoses into thermal excitations in the FL. The latter has  $S_0=0$, and hence even an arbitrary slow quench must lead to $\Delta E(\tau\to\infty)\neq 0$ and a $T\neq 0$ state, having a thermal entropy that at least accounts for the $T=0$ entropy of the initial NFL state. Hence, the observed powerlaw, implying asymptotic approach to the  adiabatic limit $\Delta E(\tau\to\infty)=0$, is surprising. It would suggest that $S_0$ is not manifested as thermal excitations in the final state for $\tau\to \infty$. Hence, we do not expect the power law to eventually persist for any finite $p$ for $\tau\to \infty$. We see  an indication of only a weak deviation from the intermediate-$\tau$ power law for $p=10$ around $\tau\sim 30-40$ (\fig{fig:SlowQuench}(a)(inset)). From the intermediate-$\tau$ power law, we can estimate a time scale, much longer than presently accessible in our calculations, where the scaling is expected to be violated due to $S_0$ (see Appendix \ref{subsec:powlawbreak}). This mechanism of the violation of adiabaticity due to $S_0$ in the large-$N$ limit is different from the hitherto known routes \cite{Polkovnikov2011} of adiabaticity breaking. As we show in Appendix \ref{subsec:powlawbreak}, physics beyond the large-$N$ limit \cite{Bagrets2016} suggests that the limits $\tau\to\infty$ and $N\to\infty$ do not commute, also indicating the absence of the adiabatic limit \cite{Polkovnikov2011}.

As shown in \fig{fig:SlowQuench}(c), \ttdn{a} steady-state value of $E_{c\psi}(t)$, consistent with the thermal value is attained very rapidly within the NFL for any $\tau$. In contrast, a `glassy' behavior is seen within the FL, where $E_{c\psi}(t)$ relaxes very slowly for small $\tau$, but relaxes almost instantaneously for larger $\tau$ values (\fig{fig:SlowQuench}(d)). 

\section{Conclusions} \label{sec:Conclusion}
In conclusion, our study of sudden and slow quenches in a large-$N$ model of NFL-FL transition reveal a remarkably rich non-equilibrium phase diagram and sharp contrasts between non-interacting, FL and NFL phases. 
 The sudden quench allows us to track the distinct evolutions of initially prepared well defined quasipartcile state in the NFL and FL phases and establish the existence of a dynamical phase transition which is different from the equilibrium NFL-FL quantum phase transition.  In the context of slow quenches, unique features of the NFL-FL QPT and the low-temperature state of SYK model, such as strongly interacting fermionic excitations and residual zero-temperature entropy, allows us to probe completely unexplored regime of out-of-equilibrium quantum many-body dynamics compared to previous studies of integrable and weakly-integrable systems. These unusual features lead to remarkable intermediate non-analytic scaling of excitation-energy production with the quench-duration and the eventual breakdown of quantum adiabaticity. A natural future extension would be to go beyond large-$N$ to study evolution for longer times $\sim N$.


\begin{acknowledgements}
We thank Ehud Altman, Subroto Mukerjee, Diptiman Sen, Joel Moore, Vijay B. Shenoy, Emil Uzbashyan, Soumen Bag, Renato Dantas and Paul A. McClarty for useful discussions. SB acknowledges support from The Infosys Foundation, India. 
\end{acknowledgements}

\appendix
\section{Non-equilibrium Green's functions and Kadanoff-Baym (KB) equations}
\mylabel{sec:neq_action}
\subsection{Interacting model}
We find the non-equilibrium Green's functions and the corresponding Kadanoff-Baym equations using the Schwinger-Keldysh closed contour formalism. To do this we  write the Schwinger-Keldysh action\mycite{KamenevBook} for the time-dependent Hamiltonian $\cH$ in \eqn{eq.Model}, i.e.
\begin{widetext}
\begin{align}\mylabel{eqn:neq_S_GBA}
\begin{array}{clc}
\action=&\ctimeint z\left[\sum\limits_{i=1}^{\Lc}\cb_{i}(z)(\ci\partial_{z}+\mu)c_{i}(z)
+\sum\limits_{\alpha=1}^{\LPsi}\Psib_{\alpha}(z)(\ci\partial_{z}+\mu)\psi_{\alpha}(z)
\right.
-\frac{1}{(2\Lc)^{3/2}}\sum\limits_{\mathclap{ijkl}}
J_{ijkl}\cb_{i}\cb_{j}c_{k}c_{l}-
{1\over \LPsi^{1/2}}\sum\limits_{\mathclap{\alpha\beta}}
 t_{\alpha\beta}^\psi\Psib_{\alpha} \psi_{\beta}\notag\\
&-\left.
{f(z)\over (\Lc\LPsi)^{1/4}}
\sum\limits_{\mathclap{i\alpha_1}}
 V_{i\alpha}\cb_{i}\psi_{\alpha}+
 V_{i\alpha}^*\Psib_{\alpha}c_{i}\right],
\end{array}\\
\end{align}
\end{widetext}
 The contour variable $z$ lies on the usual Keldysh contour \cite{KamenevBook} (\fig{fig:ctc1}) with the forward ($+$) or backward ($-$) branches. 
\begin{figure}[!h]
\centering
\includegraphics[scale=0.9]{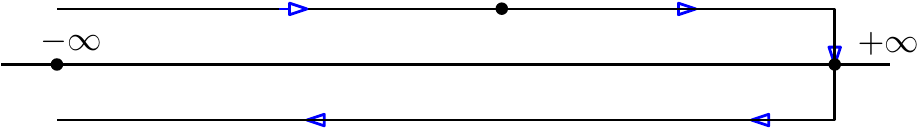} 
\caption{{\bf Schwinger-Keldysh closed time contour:} The closed-time Schwinger-Keldysh contour stretching from $-\infty$ to $+\infty$ forward in time and then backward from $+\infty$ to $-\infty$.}\mylabel{fig:ctc1}
\end{figure}
The explicit time dependence in the action is introduced via the function $f(z)=f(t)$.
The non-equilibrium generating functional for obtaining time-dependent expectation values is defined as $\Zneq=\int \mathcal{D}(c,\cb,\Psib,\psi)\ e^{\ci \action[\cb,c,\Psib,\psi]}$,
under the two usual assumptions. First, the initial density-matrix $\hat{\rho}$ is independent of any disorder, and all the disorder dependence has been pushed into the time evolution operators. Second, the disorder is switched on sometime in the infinitely long past so that the system has enough time to equilibrate to the conditions created by the disorder dependent Hamiltonian. These assumptions allow us to implement the averaging of $\Zneq$ over all disorder realizations as follows
\begin{align}
\langle \Zneq\rangle_{dis}&=\int \mathcal{D}(c,\cb,\Psib,\psi)\ \langle e^{\ci \action[\cb,c,\Psib,\psi]}\rangle_{dis}\nonumber \\
&=\int \mathcal{D}(c,\cb,\Psib,\psi)\ \int d[J,V,t]\ e^{\ci \action}P[J]P[V]P[t],\mylabel{eqn:neq_dis_avg_exp}
\end{align}
where $P[.]$ s are the Gaussian probability distributions for the couplings $J_{ijkl},~t_{\alpha\beta}^\psi$ and $V_{i\alpha}$ appearing in \eqn{eq.Model}. We perform the Gaussian integrals over the disorder distributions and define 
the large-$N$ fields,
\begin{align}
\Gc(z_{1},z_{2}) &= -\frac{\ci}{\Lc}\sum_{i}c_{i}(z_{1})\cb_{i}(z_{2}) \nonumber  \\ 
\GPsi(z_{1},z_{2})& = -\frac{\ci}{\LPsi}\sum_{\alpha}\psi_{\alpha}(z_{1})\Psib_{\alpha}(z_{2})\mylabel{eqn:GcPsi_cont}
\end{align}
that live on the contour, and the corresponding Lagrange multipliers $\Sigma_{c,\psi}(z_1,z_2)$. 
Finally, after integrating out the fermions we end up with the action
\begin{widetext}
\begin{align}
\action[\Sigma,G]=&
-\ci \Lc\ln\det\left[-\ci((\ci\partial_1+\mu)\mathbf{1}-\mathbf{\Sigc})\right]-\ci \LPsi\ln\det\left[-\ci((\ci\partial_1+\mu)\mathbf{1}-
\mathbf{\SigPsi})\right]
\nonumber\\
&\int_{\mathcal{C}}\dz_{1}\dz_{2}\Big[+\ci\frac{J^{2}\Lc}{4}\Gc(z_{2},z_{1})^{2}\Gc(z_{1},z_{2})^{2}
+\ci\frac{t_\psi^{2}\LPsi f(z_{1})f(z_{2})}{2}\GPsi(z_{2},z_{1})\GPsi(z_{1},z_{2})
\nonumber\\
&+\ci{V^{2}\sqrt{\Lc\LPsi}f(z_{1})f(z_{2})}
\Gc(z_{2},z_{1})\GPsi(z_{1},z_{2})
+\Sigc(z_{1},z_{2})[-\ci \Lc \Gc(z_{2},z_{1})]\nonumber\\
&+\SigPsi(z_{1},z_{2})[-\ci \LPsi\GPsi(z_{2},z_{1})]\Big],
\end{align}
\end{widetext}
where the matrix $(\ci\partial_1+\mu)\mathbf{1}$ has elements of the form $(\ci\partial_{z_{1}}+\mu)\delta(z_{1}-z_{2})$. The elements for matrices $\mathbf{\Sigc}$, $\mathbf{\SigPsi}$ are $\Sigc(z_{1},z_{2})$, $\SigPsi(z_{1},z_{2})$ respectively.
The action $\action$ is extremized with respect to $G$ and $\Sigma$ to produce the large-$N$ saddle point equations
\begin{align}\mylabel{eq:sigma_from_G_contour_SYKq}
\Sigc(z_1,z_2) & = 
J^2 \Gc(z_{1},z_{2})^{2}\Gc(z_{2},z_{1})\nonumber \\
& +\sqrt{p} V^2f(z_{1})f(z_{2}) 
\GPsi(z_{1},z_{2})\notag\\
 \SigPsi(z_1,z_2) & = 
t_\psi^2\GPsi(z_{1},z_{2})
+
\frac{V^2}{\sqrt{p}}f(z_{1})f(z_{2}) 
\Gc(z_{1},z_{2})
\end{align}
and
\begin{align}
(\ci\partial_{z_{1}}+\mu)\delta(z_{1}-z_{2})-\Sigma_s(z_{1},z_{2})=&G_s^{-1}(z_{1},z_{2}),\label{eq:Ginverse_on_contour}
\end{align}
where $G_s^{-1}$ is the inverse of the matrix $G_s$ with its elements given by $G_s(z_1,z_2)$ and $s=c,\psi$. We rewrite \eqn{eq:Ginverse_on_contour} by multiplying with $G_s$ from the right and the left, respectively,
\begin{align}
(\ci\partial_{z_{1}}+\mu)G_s(z_{1},z_{2})-&\int_{\mathcal{C}}\dz\Sigma_s(z_{1},z)G_s(z,z_{2})\nonumber \\
&=\delta_{\mathcal{C}}(z_{1},z_{2}),\mylabel{eq:KB_eqn_cont_1}\\
\left(-\ci\partial_{z_{2}}+\mu\right)G_s(z_{1},z_{2})-&\int_{\mathcal{C}}\dz\ G_s(z_{1},z)\Sigma_s(z,z_{2})\nonumber \\
&=\delta_{\mathcal{C}}(z_{1},z_{2})\mylabel{eq:KB_eqn_contour_2},
\end{align}
which are integro-differential equations satisfied by $G_s(z_1,z_2)$, and where $\delta_c$ is the Dirac-delta function defined on the contour. 

From the contour-ordered Green's function $G_s(z_1,z_2)$, we obtain the disorder averaged real-time non-equilibrium Green's functions, greater ($>$), lesser ($<$), and retarded $R$, from $G_s(z_1,z_2)$ at the saddle point, e.g.
\begin{align}
G_c^{>}(t_{1},t_{2})&\equiv G_c(t_{1-},t_{2+})=-\ci\overline{\langle c_i(t_{1})c_i^{\dagger}(t_{2})\rangle} \label{eq:G>},\\
G^{<}_c(t_{1},t_{2})&\equiv G_c(t_{1+},t_{2-})=+\ci\overline{\langle c_i^{\dagger}(t_{2})c_i(t_{1})\rangle} \label{eq:G<},\\
G^R(t_{1},t_{2})&=-\ci\Theta(t_1-t_2)\overline{\langle \{c_i(t_{1}),c_i^{\dagger}(t_{2})\}\rangle}\nonumber \\
&=\Theta(t_{1}-t_{2})\left[G^{>}(t_{1},t_{2})-G^{<}(t_{1},t_{2})\right],\label{eq:GR}\\
G^A(t_{1},t_{2})&=+\ci\Theta(t_2-t_1)\overline{\langle \{c_i(t_{1}),c_i^{\dagger}(t_{2})\}\rangle}\nonumber \\
&=\Theta(t_{2}-t_{1})\left[G^{<}(t_{1},t_{2})-G^{>}(t_{1},t_{2})\right].\label{eq:GA}
\end{align} 
and similarly for $G_\psi(z_1,z_2)$. The first (second) sign in the suffix of $G_s(t_{1\pm},t_{2\mp})$ indicates whether the $z_1$($z_2$) coordinate lies in the forward (+ve) or backward (-ve) branch of the contour.

\emph{Steady state:} In a steady-state the Green's functions are invariant under time translational, e.g., $G^R(t_1,t_2)=G^R(t_1-t_2)=G^R(t)$.

{\it Thermal equilibrium:} For thermal equilibrium at a temperature $T$, in addition to the above steady-state condition, the Green's functions satisfy the fluctuation dissipation theorem (FDT) \mycite{KamenevBook}, e.g.
\begin{align}
\ci G_s^<(\omega)/(2\mr{Im}G_s^R(\omega))&=n_\mr{F}(\omega,T),\mylabel{eqn:FDT}
\end{align}
where $n_\mr{F}(\omega,T)=1/(e^{\omega/T}+1)$ is the Fermi function
and $G_{<,>,R,A}(\omega)$ are Fourier transforms of $G_{<,>,R,A}(t)$ defined by
\bea
G(\omega)=\int_{-\infty}^{\infty} G(t)e^{\ci\omega t}\dw,
\eea
The above conditions allow us to test whether a system, after undergoing a non-equilibrium process, has reached a steady state or not, and whether the steady state is consistent with thermal equilibrium.

{\bf The Kadanoff-Baym Equations}\mylabel{sec:KBeqns}
From \eqn{eq:KB_eqn_cont_1} and \eqn{eq:KB_eqn_contour_2}, we obtain the time-evolution equations for $G_s^{>,<}$, e.g. 
\begin{eqnarray}
(\ci\partial_{t_{1}}+\mu)G_s^>(t_{1},t_{2}) & = &\int_{\mathcal{C}}\dz\Sigma_s(t_{1-},z)G_s(z,t_{2+})\nonumber \\
&\equiv& I^{(1)}_{s>}(t_1,t_2)\nonumber \\
(-\ci\partial_{t_{2}}+\mu)G^>_s(t_{1},t_{2}) & = &\int_{\mathcal{C}}\dz\ G_s(t_{1-},z)\Sigma_s(z,t_{2+})\nonumber \\
&\equiv& I^{(2)}_{s>}(t_1,t_2)\mylabel{eq:KB_eqn_contour_2a},
\end{eqnarray}
where we have used the fact $\delta_\mathcal{C}(z_{1}\to t_{1-},z_{2}\to t_{2+})=0$. 
Finally, using Langreth rules\mycite{langrethPRB1972} we get
\begin{align}
I^{(1)}_{s>}(t_1,t_2) & = \int_{-\infty}^{t_{1}}\Sigma_s^{R}(t_{1},t)G_s^{>}(t,t_{2})\dt\nonumber \\
&+\int_{-\infty}^{t_{2}}\Sigma_s^{>}(t_{1},t)G_s^{A}(t,t_{2})\dt\nonumber \\
I^{(2)}_{s>}(t_1,t_2) & = \int_{-\infty}^{t_{1}}G_s^{R}(t_{1},t)\Sigma^{>}(t,t_{2})\dt\nonumber \\
&+\int_{-\infty}^{t_{2}}G_s^{>}(t_{1},t)\Sigma_s^{A}(t,t_{2})\dt\label{eq:I_KB_eq_1},
\end{align}
where $\Sigma^{R}$ ($\Sigma^{A}$) is the retarded (advanced) self energies, given by
\begin{align}\mylabel{eqn:SR_SA_def}
\Sigma^{R}(t_{1},t_{2}) & = \Theta(t_{1}-t_{2})[\Sigma^{>}(t_{1},t_{2})-\Sigma^{<}(t_{1},t_{2})]\nonumber \\
\Sigma^{A}(t_{1},t_{2}) & = \Theta(t_{2}-t_{1})[\Sigma^{<}(t_{1},t_{2})-\Sigma^{>}(t_{1},t_{2})], \
\end{align}
where, using \eqn{eq:sigma_from_G_contour_SYKq},
\begin{align}\mylabel{eqn:Sg_Sl_def}
&\Sigma_c^{>,<}(t_1,t_2)=\Sigma_c(z_1\to t_{1-},z_2\to t_{2+})\nonumber\\
&=
J^2G^{>,<}_c(t_{1},t_{2})^{2}G^{<,>}_c(t_{2},t_{1})+
\sqrt{p}\ V^2f(t_{1})f(t_{2}) 
G^{>,<}_{\psi}(t_{1},t_{2})
\nonumber\\
&\Sigma_\psi^{>,<}(t_1,t_2)=\Sigma_\psi(z_1\to t_{1-},z_2\to t_{2+})
=
t_\psi^2G^{>,<}_\psi(t_{1},t_{2})\nonumber \\
 &+
{V^2\over\sqrt{p}} f(t_{1})f(t_{2}) 
G^{>,<}_{c}(t_{1},t_{2}).
\end{align}
Similarly, one can repeat the above analysis for the $G_s^<$ case.
The integro-differential equations, for $G_s^>$ and $G_s^<$ can be stated together as
\begin{align}\label{eq:KB_eq_1_sec}
(\ci\partial_{t_{1}}+\mu)G_s^{>,<}(t_{1},t_{2}) & =  \int_{-\infty}^{t_{1}}\Sigma_s^{R}(t_{1},t)G_s^{>,<}(t,t_{2})\dt\nonumber \\
&+\int_{-\infty}^{t_{2}}\Sigma_s^{>,<}(t_{1},t)G_s^{A}(t,t_{2})\dt\nonumber \\
(-\ci\partial_{t_{2}}+\mu)G_s^{>,<}(t_{1},t_{2}) & =  \int_{-\infty}^{t_{1}}G_s^{R}(t_{1},t)\Sigma_s^{>,<}(t,t_{2})\dt\nonumber \\
&+\int_{-\infty}^{t_{2}}G_s^{>,<}(t_{1},t)\Sigma_s^{A}(t,t_{2})\dt.
\end{align}
This set of equations are called the \emph{Kadanoff-Baym} (KB) equations, which along with the relations given in \eqn{eqn:SR_SA_def} and \eqn{eqn:Sg_Sl_def}, set up a closed system of equations that can be time evolved in the $t_1-t_2$ plane (see 
\figSM{fig:KB_eqn_t1t2_palne_app}), starting from an initial condition for $G$s and $\Sigma$s.

\subsection{Non-interacting model}
\mylabel{sec:nI_model}
Following procedure similar to that discussed above for the interacting model, we obtain the disorder averaged Schwinger-Keldysh action for the non-interacting model of \eqref{eq.SYK_SI} and the saddle point equations.

The large-$N$ self-energies are given by
\begin{align}\mylabel{eq:sigma_from_G_contour_SYKq_nI}
\Sigc(z_1,z_2) & = 
t_c^2 \Gc(z_{1},z_{2})
 +
\sqrt{p} V^2f(z_{1})f(z_{2}) 
\GPsi(z_{1},z_{2})\notag\\
 \SigPsi(z_1,z_2) & = 
t_\psi^2\GPsi(z_{1},z_{2})
+
\frac{V^2}{\sqrt{p}}f(z_{1})f(z_{2}) 
\Gc(z_{1},z_{2})
\end{align}
Consequently, a set of contour Kadanoff-Baym equations similar to the ones given in \eqn{eq:KB_eqn_cont_1} and \eqn{eq:KB_eqn_contour_2} is obtained. Using the Langreth rules the above equations involving contour indices $z_1$,$z_2$ can be changed to real time variables $t_1$, $t_2$ to give the final Kadanoff-Baym equations (see \eqn{eq:KB_eq_1_sec}) for the system
with the following expressions for the self-energies
\begin{align}\mylabel{eqn:Sg_Sl_nI_def}
\Sigma_c^{>,<}(t_1,t_2)
=&
t_c^2G^{>,<}_c(t_{1},t_{2})
 +
\sqrt{p}\ V^2f(t_{1})f(t_{2}) 
G^{>,<}_{\psi}(t_{1},t_{2})
\nonumber\\
\Sigma_\psi^{>,<}(t_1,t_2)
=&
t_\psi^2G^{>,<}_\psi(t_{1},t_{2})
 +
{V^2\over\sqrt{p}} f(t_{1})f(t_{2}) 
G^{>,<}_{c}(t_{1},t_{2}).
\end{align}
\section{Gap closing transitions in the interacting model and non-interacting model}
\begin{figure*}
\includegraphics[scale=0.8]{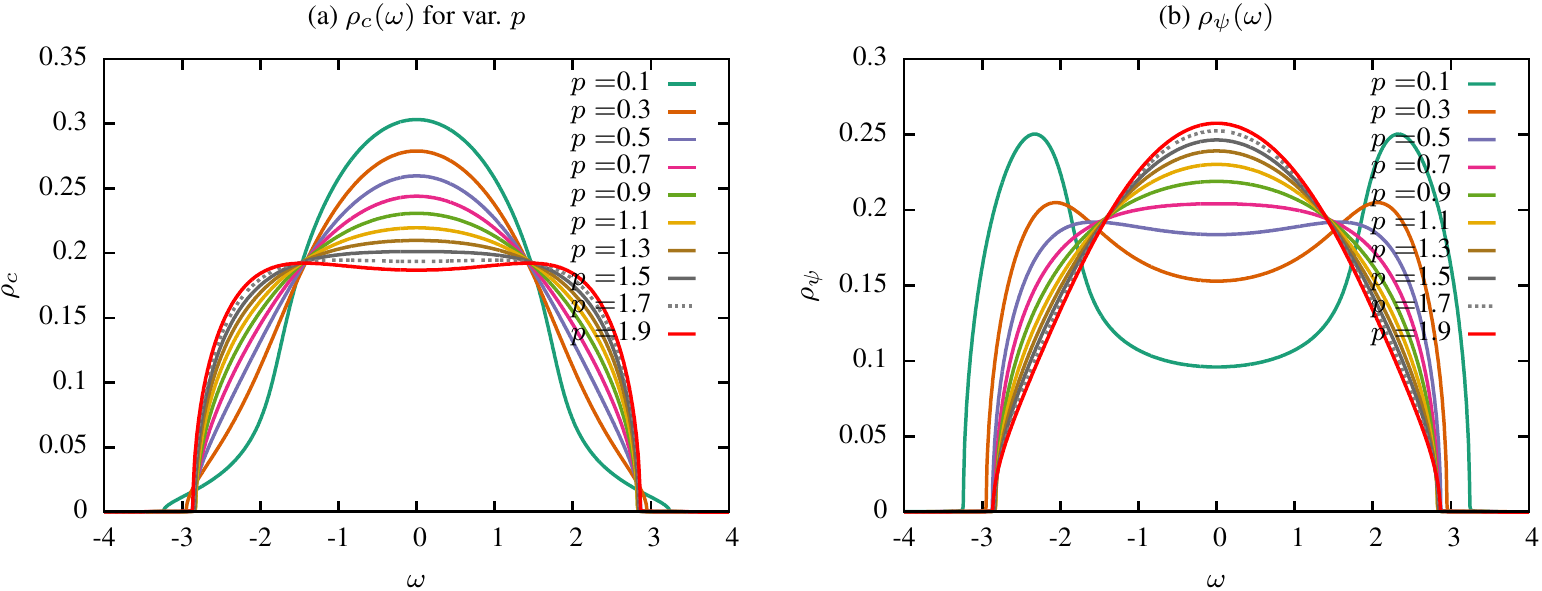} 
\caption{{\bf Spectral functions for the non-interacting case at $T=0.04$:} (a) The spectral functions for the $c$-fermions shown for multiple values of $p$. (b) Spectral function for the $\psi$-fermions, showing the soft-gap closing between $p=0.5$ and $p=0.7$. 
}\label{fig:q11_spectral_func}
\end{figure*}

\begin{figure*}
\includegraphics[scale=0.8]{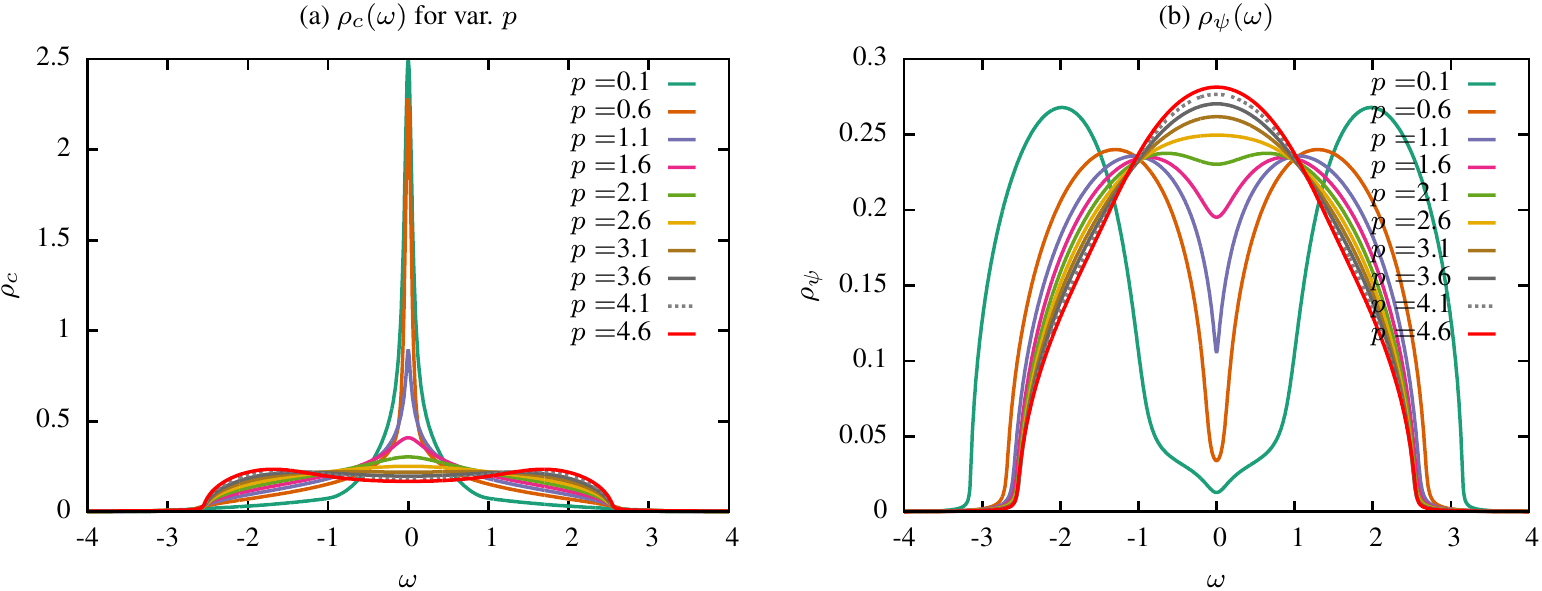} 
\caption{{\bf Spectral functions for the interacting model at $T=0.04$:} (a) Spectral function for the $c$-fermions for multiple values of $p$ . (b) Spectral function for the $\psi$-fermions, showing the soft-gap closing somewhere between $p=1.6$ and $p=2.6$
.}\label{fig:q21_spectral_func}
\end{figure*}

\mylabel{sec:gap_closing}
The spectral functions obtained from the equilibrium Green's functions (see 
\SMcite{sec:eq_Gfs}) for the connected system, i.e. $V=1$, for various $p$-values are discussed below. \Fig{fig:q11_spectral_func}(a) and (b) show the results for the non-interacting model, whereas \fig{fig:q21_spectral_func}(a) and (b) show the spectral functions for the interacting model. In the non-interacting case, we find that the spectral function for the $\psi$ fermions have a soft-gap for smaller values of $p$, which closes completely around $p=0.5-0.7$. This is consistent with the dynamical transition that we observe for the sudden quench of the non-interacting model, which we discuss in 
\SMcite{sec:nI_sudden_results}. The $c$-fermion spectral function does not have a soft-gap for small values of $p$, but a soft-gap begins to form near $p\sim 1.9$ as we increase $p$. This is expected since, the non-interacting model has an additional symmetry under $p\to 1/p$ and $c\leftrightarrow \psi$. Moving on to the interacting model, we find a similar soft-gap closing scenario taking place for the $\psi$-fermion spectral functions, as shown in \fig{fig:q21_spectral_func}(b). However, this time the gap closes completely around $p=1.6-2.6$, which is far away from the equilibrium NFL to FL transition point of $p=1.0$. This is again consistent with the dynamical transition critical point (see \fig{fig:SuddenQuench}(a) inset in the main text) that we find in the sudden quench of the interacting model. The $c$-fermion spectral functions have a highly peaked form around $\omega=0$, for smaller $p$ values, due to the presence of a divergent $T=0$ spectral function coming from the NFL fixed point. At higher values of $p$, the peak subsides and a gap begins to form in the spectral function.

\section{Excitation energy as a function of quench time for the non-interacting model}\mylabel{subsec:excnI}

\begin{figure}[!h]
\includegraphics[scale=1]{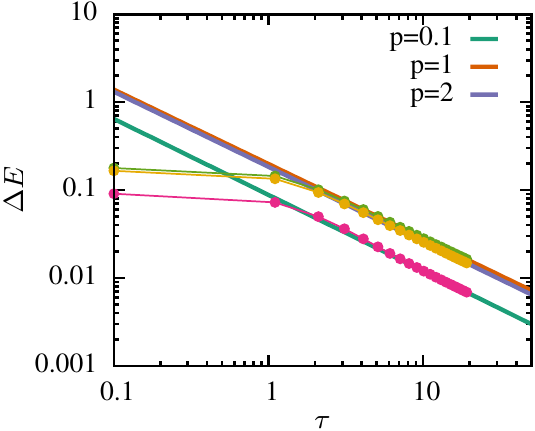} 
\caption{{\bf Excitation energy vs. quench duration for the non-interacting model:} The excitation energy $\Delta E$, produced in the slow quench, plotted as a function of quench duration $\tau$. The powerlaw dependence is clearly evident from the straight line fits to the log-log scale plot of $\Delta E$ vs. $\tau$. The slope of the straight lines are almost equal making the exponent $\eta$ of the powerlaw to have only a weak dependence on the site fraction $p$. This feature is in contrast with the $\eta-p$ dependence for the interacting model, where $\eta$ heavily depends on $p$ and has a minimum value at $p=1$, see \fig{fig:SlowQuench}(b).}\label{fig:powlaw_nI}
\end{figure}
We also study the dependence of excitation energy $\Delta E$ on the quench duration $\tau$ for the slow quench of the non-interacting model of \eqn{eq.SYK_SI}. We find that there exists a powerlaw relationship between $\Delta E$ and $\tau$ here as well, as shown in \fig{fig:powlaw_nI}, which we report in the main text. However, the dependence of the powerlaw exponent $\eta$ on $p$ is qualitatively different from the interacting case as shown in \fig{fig:SlowQuench}(b) of the main text. 

\section{Breakdown of adiabatic perturbation theory and the absence of adibatic limit}\mylabel{subsec:powlawbreak}
Here we elaborate on the connection between the zero-temperature residual entropy of the SYK NFL and the absence of the adiabatic limit for the slow quenches described in the main paper. In the large-$N$ limit, $S_0$ reflects the exponentially dense many-body energy spectrum near the ground state in the NFL phase \cite{Maldacena2016}, and the QPT in our model marks a transition from exponentially small many-body level spacing, $\Delta\sim e^{-S_0(p)N}$, in the NFL to $\Delta\sim 1/N$ in the FL. Hence, the residual entropy $S_0$ cannot be thought of as a thermodynamic entropy strictly at $T=0$, i.e., when the $T\to 0$ limit is taken first, keeping $N$ finite and then $N\to\infty$ limit is taken, $S_0=0$. In the large-$N$ description, the limit is taken the other way around, and it captures the exponentially dense many-body level spectrum near the ground state in the NFL phase. However, at any non-zero temperature ($\apgt e^{-S_0N}$), which could be infinitesimally small in the large-$N$ limit, $S_0$ is the true thermodynamic entropy. Hence, we expect this entropy to be manifested in the large-$N$ non-equilibrium dynamics during a slow quench, implying the absence of the adiabatic limit. However, as mentioned in the main paper, surprisingly we find the intermediate-$\tau$ non-analytic power law scaling, that seems to mask the effect of residual entropy. In the following, we first try to estimate the power law from the so-called adiabatic perturbation theory \cite{Polkovnikov2005,Dziarmaga2010,DeGrandi2010,Eckstein2010}, that has been previously used for non-interacting and weakly-interacting systems. We show that the adibatic perturbation theory cannot explain the $p$-dependent exponent directly obtained from the direct non-equilibrium evolution, discussed in the main text. Furthermore, we discuss the possible modes of violation of the adiabaticity in the large-$N$ limit, as well as, beyond it.

\subsubsection{Adiabatic perturbation theory in the large-$N$ limit}
In the adiabatic perturbation theory \cite{Polkovnikov2005,Dziarmaga2010,DeGrandi2010,Eckstein2010}, the time-dependent term (eqn.\eqref{eq.Coupling}) is treated as a perturbation assuming a weak strength of the ramp in eqn.\eqref{eq.Model}, i.e. $f(t)=\Delta f r(t/\tau)$ with $\Delta f\ll 1$. To this end, we obtain the energy generated during quench \cite{Eckstein2010} as,
\begin{align}
\Delta E(\tau)&\simeq \Delta f^2\int_0^\infty\frac{d\omega}{\omega}\mathcal{A}_\mathcal{V}(\omega)R(\omega\tau)+\mathcal{O}(\Delta f^2). 
\end{align}
Here $R(x)=|\int_0^1ds r'(s)e^{ixs}|^2$, $r'(x)$ is derivative of the ramp function. $\mathcal{A}_\mathcal{V}(\omega)$ is a $T=0$ spectral function corresponding to the disordered-averaged imaginary-time correlation function of the operator $\mathcal{V}=(NM)^{-1/4}\sum_{i\alpha}(V_{i\alpha}c_i^\dagger \psi_\alpha+\mr{h.c.})$, i.e. $-\overline{\langle \mathcal{T}_\tau \mathcal{V}(\tau)\mathcal{V}(0)\rangle}$. This can be computed in the large-$N$ limit, and we obtain,
\begin{align}
\mc{A}_\mc{V}(\omega)&=\frac{2\sqrt{p}}{1+p}V^2\int_0^\omega d\omega'\rho_c(\omega')\rho_\psi(\omega'-\omega),
\end{align}
where $\rho_c(\omega)$ and $\rho_\psi(\omega)$ are the spectral functions of the SYK and lead fermions, respectively, for the uncoupled systems before the quench. For long quench time $\tau\gg \Omega^{-1}$, where $\Omega\approx J,t_\psi$, we can use the low-energy forms, $\rho_c(\omega)\sim |\omega|^{-1/2}$ and $\rho_\psi(\omega)\sim$ constant, for $\omega\ll \Omega$. These give $\mc{A}_\mc{V}(\omega)\sim \omega^{1/2}$. It can be shown \cite{Eckstein2010}, $R(x\to\infty)\sim x^{-2n}$, where $n\geq 1$ is the order of the derivative discontinuity in the ramp (see 
\SMcite{subsec:rampshapeseffect}). As a result,
\begin{align}
\Delta E(\tau)\sim \tau^{-1/2}\int_0^{\Omega \tau} dxx^{-1/2}R(x).
\end{align}
Moreover, the integral above is convergent for $\tau \to \infty$ since $2n>1/2$. Hence, the adiabatic perturbation theory predicts a non-analytic power law with exponent $\eta=1/2$ for any $p$. This, of course, does not agree with the strongly non-monotonic $\eta(p)$ obtained from the direct non-equilibrium calculations (fig.\ref{fig:SlowQuench}(b)). Hence, the adiabatic perturbation theory does not work in our case. The theory assumes non-degenerate ground state \cite{Eckstein2010}, and it is an interesting question whether such theory could at all be applied to a phase with exponentially dense many-body spectrum near ground state, even though some of the effect of the dense spectrum is incorporated in the large-$N$ single-particle density of states.
\subsubsection{Breakdown of adiabaticity in the large-$N$ limit}
As discussed in the main-text, the presence of a residual entropy in the SYK fermions, prior to the quench, implies that a finite amount of excitation energy $\Delta E>0$ must be produced even in the $\tau\to\infty$ limit. Therefore, the powerlaw relationship $\Delta E \sim \tau^{-\eta}$, in principle, should break down at some large $\tau$. We now provide a route to find an estimate for the time $\tau_{\textup{break}}$ at which we can expect the powerlaw behavior to break down. This can be done by assuming an iso-entropic (the initial and final entropies are taken to be equal) limit of the quench process. We first calculate the temperature $T_{f}$ necessary for the final Hamiltonian $H_f$ to hold the initial entropy $S_i$ by solving the equilibrium problem and demanding
\begin{align}
S_i=S_f(T_f).
\end{align}
Using the definition of $\Delta E$ given in the main text
 we can estimate $\tau_{\textup{break}}$ from the condition, $\Delta E(\tau_\mr{break})\approx \Delta E_{T_f}=\langle \mc{H}(\tau_\mr{break})\rangle_{T_{f}}-\langle \mc{H}(\tau_\mr{break})\rangle_{T_i}$. The latter is the excitation energy produced by converting the initial entropy to thermal excitations, and the excitations generated by the quench cannot be lower than $\Delta E_{T_f}$. This leads to,
\begin{align}
\tau_{\textup{break}}(p)=\left[\frac{\Tr[\hat{\rho}(T_f) H_f]-\Tr[\hat{\rho}(T_i)H_f]}{\alpha}\right]^{-1/\eta},
\end{align}
where $\alpha$ can be extracted from powerlaw fits to $\Delta E$ using
\begin{align}
\Delta E=\alpha \tau^{-\eta}.
\end{align}
Performing such an estimate for $p=1.5$ case, we find $\tau_{\textup{break}}\sim 200$, a time which is not easily accessible using the numerical algorithm of 
\SMcite{sec:pred_corr_KB}.
\subsubsection{The adiabatic perturbation theory beyond large-$N$ and the absence of adiabatic limit}
Within the adiabatic perturbation theory discussed above, we can go beyond the large-$N$ theory by incorporating some finite-$N$ corrections, at least due to the single-particle level spacing, following ref.\onlinecite{Bagrets2016}. It is still not known how to incorporate the effects of many-body level spacing. Nevertheless, it has been shown in ref.\onlinecite{Bagrets2016} that the SYK spectral function $\rho_c(\omega)$ changes from the divergent $|\omega|^{-1/2}$ behavior to $\rho_c(\omega)\sim \Delta_s^{-1}|\omega|^{1/2}$ for $\omega\ll \Delta_s\sim J/(N\ln{N})$. The large prefactor $N\ln{N}$ in the $\sqrt{|\omega|}$ dependence presumably arises from the dense spectrum, even though the density of states is suppressed at low energies. We obtain $\mc{A}_\mc{V}(\omega)\sim \omega ^{3/2}$ for $\omega\ll \Delta_s$, giving
\begin{align}
\Delta E(\tau)&\sim \tau^{-3/2}N\ln{N}\int_0^{\Delta_s \tau} dxx^{1/2}R(x)\nonumber \\
&+\tau^{-1/2}\int_{\Delta_s \tau}^{\Omega \tau} dxx^{-1/2}R(x).
\end{align}
Hence, keeping $N$ fixed, we obtain $\Delta E(\tau)\sim \tau^{-3/2}N\ln{N}$ for $\tau \to \infty$. Clearly, the limits $\tau \to \infty$ and $N\to \infty$ do not commute. This indicates that the adiabatic limit can not be reached.

\bibliography{SYK}

\ifdefined\makeSM
\clearpage
\newpage

\renewcommand{\appendixname}{}
\renewcommand{\thesection}{{S\arabic{section}}}
\renewcommand{\theequation}{\thesection.\arabic{equation}}
\renewcommand{\thefigure}{S\arabic{figure}}
\setcounter{page}{1}
\setcounter{figure}{0}
\setcounter{section}{0}

\widetext

\centerline{\bf Supplemental Material}
\centerline{\bf for}
\centerline{\bf \titlename}
\centerline{by \authornames}
\affiliation{\affiliations}

%
%

\section{Numerical integration of KB equations: Predictor corrector algorithm}
\mylabel{sec:pred_corr_KB}
The time propagation of KB equations (see \eqn{eq:KB_eq_1_sec}) is done numerically on the $t_1$-$t_2$ (see \fig{fig:KB_eqn_t1t2_palne_app}) time plane using a predictor-corrector algorithm. The algorithm works by guessing a value for the Green's function ($G^<(t_1,t_2)$, $G^<(t_2,t_1)$) at a new time step by using a \emph{predictor} scheme and then uses the said value along with past values to correct the guess by using a \emph{corrector} scheme. The initial conditions for the Green's functions are encoded in quadrant $C$ as shown in \fig{fig:KB_eqn_t1t2_palne_app} using the equilibrium solutions for the functions obtained by using the initial Hamiltonian from before the quench. We have discussed the details of constructing the equilibrium Green's functions in \seccite{sec:eq_Gfs}. Before we discuss the rest of the algorithm, we define a new center of time variable $\coT$ and a relative time variable $\rT$ as 
\bea
\coT=&(t_1+t_2)/2\notag\\
\rT=&t_1-t_2,
\eea
Since,
$
\frac{\partial}{\partial\coT}=
\frac{\partial}{\partial t_1}
\frac{\partial t_1}{\partial \coT}
+
\frac{\partial}{\partial t_2}
\frac{\partial t_2}{\partial \coT}
=
\frac{\partial}{\partial t_1}
+
\frac{\partial}{\partial t_2},
$
we derive yet another differential equation in the variable $\coT$ by subtracting the two equations in \eqn{eq:KB_eq_1_sec}, and get the following
\begin{eqnarray}
(\ci\underbrace{(\partial_{t_{1}}+\partial_{t_{2}})}_{\partial_ \coT})G^{>,<}(t_{1},t_{2}) & = & \int_{-\infty}^{t_{1}}\Sigma^{R}(t_{1},t)G^{>,<}(t,t_{2})\dt+\int_{-\infty}^{t_{2}}\Sigma^{>,<}(t_{1},t)G^{A}(t,t_{2})\dt\nonumber \\
&  &- \int_{-\infty}^{t_{1}}G^{R}(t_{1},t)\Sigma^{>,<}(t,t_{2})\dt
-
\int_{-\infty}^{t_{2}}G^{>,<}(t_{1},t)\Sigma^{A}(t,t_{2})\dt,\notag\\
\label{eq:KB_eq_diag_1_app}
\end{eqnarray}
which we shall use later on to time step the Green's functions when $t_1=t_2$. We now discuss the two parts of the algorithm-- part a) is the predictor scheme, which we implement using Euler discretization, and part b) a corrector scheme which will be implemented implicitly.
\begin{figure}[h!]
\centering
\ifdefined\showfigures
\includegraphics[scale=0.8]{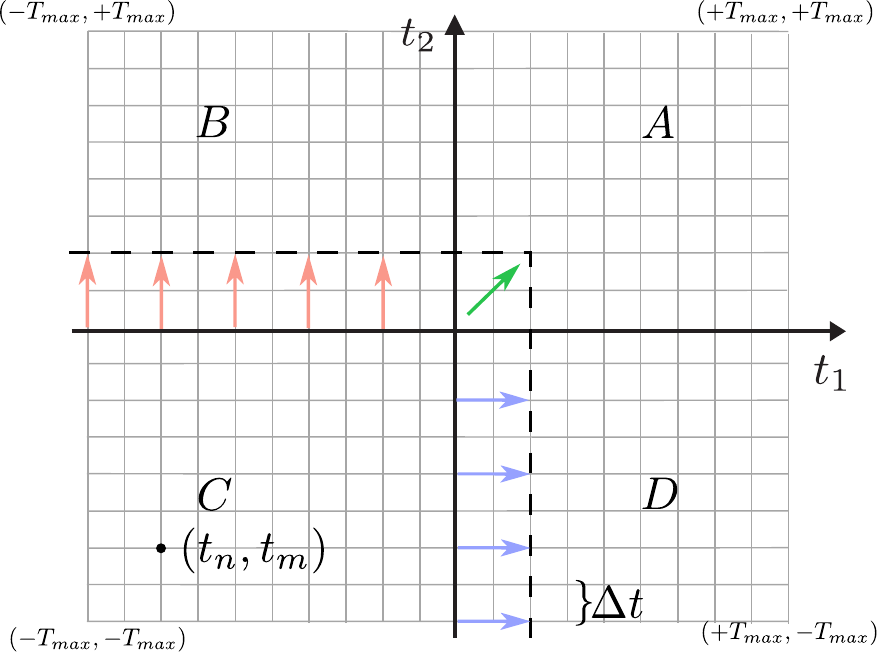}
\fi
\caption{{\bf Time evolving the Kadanoff-Baym equations:} The $t_1-t_2$ plane on which the KB equations are time evolved to obtain the values for the nonequilibrium Green's functions $G^>$ and $G^<$.  The time plane is discretized into squares of size $\Delta t$ and has a total length of $2 T_{max}$ along each side, stretching equally both forward and backward in time. The initial values for the Green's functions are encoded into the quadrant $C$ from which they are time evolved along the directions marked by the arrows. The $G^>$($G^<$) function values residing along the vertical (horizontal) line are time evolved in the direction of the blue(red) arrows, and then the anti-hermitian property of the said Green's function is used to evaluate the values on the horizontal(vertical) line. The equal time values for the Green's functions are time evolved separately along the direction marked by the green arrow.}\mylabel{fig:KB_eqn_t1t2_palne_app}
\end{figure}
\subsection{Predictor scheme}
We define $F_{1,2}^{>,<}$, such that

\begin{eqnarray}\mylabel{eqn:KB_app_2}\label{eq:KB_eq_integrals}.
\partial_{t_{1}}G^{>,<}(t_{1},t_{2}) &= -\ci\left[\int_{-\infty}^{t_{1}}\Sigma^{R}(t_{1},t)G^{>,<}(t,t_{2})\dt+\int_{-\infty}^{t_{2}}\Sigma^{>,<}(t_{1},t)G^{A}(t,t_{2})\dt\right] &\equiv F_{1}^{>,<}(t_{1},t_{2})\nonumber \\
\partial_{t_{2}}G^{>,<}(t_{1},t_{2}) &= +\ci\left[\int_{-\infty}^{t_{1}}G^{R}(t_{1},t)\Sigma^{>,<}(t,t_{2})\dt+\int_{-\infty}^{t_{2}}G^{>,<}(t_{1},t)\Sigma^{A}(t,t_{2})\dt\right] &\equiv F_{2}^{>,<}(t_{1},t_{2}).
\end{eqnarray}
We discretize the time domain and represent the discretized times with indexed symbols $t_{n},t_{m},t_{i}$ etc., see \fig{fig:KB_eqn_t1t2_palne_app}, also $\Delta t$ is the step size of our time domain and $-\infty$ ($\infty$) is a large negative (positive) value set by $-T_{max}$ ($T_{max}$). Using the said definitions the $F_{1,2}$ terms can be approximated as
\begin{align}\mylabel{eqn:F12_discrete}
F_{1}^{>,<}(t_{n},t_{m})  = & -\ci\Delta t\left[\sum_{i=-\infty}^{n}\Sigma^{R}(t_{n},t_{i})G^{>,<}(t_{i},t_{m})+\sum_{i=-\infty}^{m}\Sigma^{>,<}(t_{n},t_{i})G^{A}(t_{i},t_{m})\right]\nonumber \\
F_{2}^{>,<}(t_{n},t_{m})  = & +i\Delta t\left[\sum_{i=-\infty}^{n}G^{R}(t_{n},t_{i})\Sigma^{>,<}(t_{i},t_{m})+\sum_{i=-\infty}^{m}G^{>,<}(t_{n},t_{i})\Sigma^{A}(t_{i},t_{m})\right].
\end{align}
The Green's functions at $(t_{n+1}=t_{n}+\Delta t,t_{m})$ and $(t_{n},t_{m+1})$ can then be predicated using
\begin{eqnarray}
G^{>,<}(t_{n+1},t_{m}) & = & G^{>,<}(t_{n},t_{m})+\Delta t F_{1}^{>,<}(t_{n},t_{m})\nonumber \\
G^{>,<}(t_{n},t_{m+1}) & = & G^{>,<}(t_{n},t_{m})+\Delta t F_{2}^{>,<}(t_{n},t_{m}),
\end{eqnarray}
in the directions indicated by the blue and red arrows in \fig{fig:KB_eqn_t1t2_palne_app}. Suppose we have all the information, i.e. $G^{>,<},\Sigma^{>,<}$,
in the grid $[-\infty,t_{k}]\times[-\infty,t_{k}]$ and want to extend
it to $t_{k+1}=t_{k}+\Delta t$, then $F_{1,2}^{>,<}$, for $n,m\leq k$, can be easily calculated since the information needed to evaluate $F_{1,2}^{>,<}$ is readily available. Although the equations in \eqn{eqn:KB_app_2} are perfectly valid for obtaining a prediction for both $G^>$, $G^<$ along the horizontal and vertical lines in \fig{fig:KB_eqn_t1t2_palne_app}, they are computationally expensive. We can cut down on the cost by using the following property of the nonequilibrium Green's functions
\begin{align}\mylabel{eqn:cc_prop_neqG}
G^{>}(t_1,t_2)^*=&-G^{>}(t_2,t_1)\notag\\
G^{<}(t_1,t_2)^*=&-G^{<}(t_2,t_1).
\end{align}
To this end, we time step one of the Green's functions, for e.g. $G^{>}$, along the vertical line in the direction of the blue arrows shown in \fig{fig:KB_eqn_t1t2_palne_app}, i.e. $(t_{n},t_{m})\to (t_{n+1},t_{m})$, and the lesser Green's function $G^{<}$ along the horizontal line in the direction of the red arrows, i.e. $(t_{n},t_{m})\to (t_{n},t_{m+1})$, and then use \eqn{eqn:cc_prop_neqG} to find $G^{>}$ ($G^{<}$) on the horizontal (vertical) line as follows
\begin{align}
G^{>}(t_{n},t_{m+1})=&-G^{>}(t_{n+1},t_{m})^*\notag\\
G^{<}(t_{n+1},t_{m})=&-G^{<}(t_{n},t_{m+1})^*.
\end{align}
Note that we have not yet obtained a prediction for the diagonal point $(t_{k+1},t_{k+1})$, i.e, $G^{>,<}(t_{k+1},t_{k+1})$. This can be done using \eqn{eq:KB_eq_diag_1_app}, which gives us the prediction formula
\begin{align}
G^{>,<}(t_{k+1},t_{k+1})=G^{>,<}(t_{k},t_{k})+
\Delta t(F_{1}^{>,<}(t_{k},t_{k})+F_{2}^{>,<}(t_{k},t_{k})).
\end{align} 

\subsection{Corrector scheme}
In order to correct the values for the Green's function at $(t_{n=k+1},t_{-\infty\leq m\leq k})$, $(t_{-\infty\leq n\leq k},t_{m=k+1})$ and $(t_{k+1},t_{k+1})$ we need to find $F_{1}^{>,<}(t_{n=k+1},t_{m})$, $F_{2}^{>,<}(t_{n},t_{m=k+1})$ and $F_{2}^{>,<}(t_{n=k+1},t_{m=k+1})$ which can be done by substituting the predicted values for $G^{>,<}$, derived earlier, into \eqn{eqn:F12_discrete} and obtain
\begin{align}
F_{1}^{>,<}(t_{n=k+1},t_{-\infty\leq m\leq k}) & =  -\ci\Delta t\left[\sum_{i=-\infty}^{n=k+1}\Sigma^{R}(t_{n=k+1},t_{i})G^{>,<}(t_{i},t_{m})
+\sum_{i=-\infty}^{-\infty\leq m\leq k}\Sigma^{>,<}(t_{n=k+1},t_{i})G^{A}(t_{i},t_{m})\right]\nonumber \\
F_{2}^{>,<}(t_{-\infty\leq n\leq k},t_{m=k+1}) & =  +i\Delta t\left[\sum_{i=-\infty}^{-\infty\leq n\leq k}G^{R}(t_{n},t_{i})\Sigma^{>,<}(t_{i},t_{m=k+1})
+\sum_{i=-\infty}^{m=k+1}G^{>,<}(t_{n},t_{i})\Sigma^{A}(t_{i},t_{m=k+1})\right].
\end{align}
A prediction for the self-energies appearing above can be obtained by using the closure relation given in \eqn{eqn:Sg_Sl_def}. Having calculated the new values for $F_{1,2}$, we can now get a better estimate for the Green's functions as follows
\begin{align}
G^{>,<}(t_{n=k+1},t_{-\infty\leq m\leq k}) & =  G^{>,<}(t_{n},t_{m})+\frac{\Delta t}{2}\left[F_{1}^{>,<}(t_{n=k+1},t_{-\infty\leq m\leq k})+F_{1}^{>,<}(t_{n=k},t_{-\infty\leq m\leq k})\right]\nonumber \\
G^{>,<}(t_{\infty\leq n\leq k},t_{m=k+1}) & =  G^{>,<}(t_{n},t_{m})+\frac{\Delta t}{2}\left[F_{2}^{>,<}(t_{-\infty\leq n\leq k},t_{m=k+1})+F_{2}^{>,<}(t_{-\infty\leq n\leq k},t_{m=k})\right].
\end{align}
Using these corrected values we can find the corrected value for $G^{>,<}(t_{k+1},t_{k+1})$ as well, which is
\begin{align}
G^{>,<}(t_{k+1},t_{k+1})=&G^{>,<}(t_{k},t_{k})+
{\Delta t\over 2}\left[F_{1}^{>,<}(t_{k+1},t_{k+1})+F_{2}^{>,<}(t_{k+1},t_{k+1})+ F_{1}^{>,<}(t_{k},t_{k})+F_{2}^{>,<}(t_{k},t_{k}))\right].
\end{align} 
The above correction process can be carried out arbitrary number (we have found that one corrective iteration is sufficient to get good results) of times till one finds that the new values for $[-\infty,t_{k+1}]\cup[-\infty,t_{k+1}]$ have converged to a desired  accuracy.

\subsection{Evaluation schemes for the integrands}
In the above discussion we chose a very simple integration scheme
for the integrals appearing in the KB equations for the sake of clarity. We now slightly modify the scheme to get a better error performance. We approximate the integrands in \eqn{eq:KB_eq_integrals} using the trapezoidal rule, such that
\begin{align}
F_{1}^{>,<}(t_{n},t_{m}) & =  -\ci\Delta t\left[\sum_{i=-\infty}^{n-1}\frac{1}{2}\left(\Sigma^{R}(t_{n},t_{i})G^{>,<}(t_{i},t_{m})+\Sigma^{R}(t_{n},t_{i+1})G^{>,<}(t_{i+1},t_{m})\right)\right.
\notag\\
&
\left.
+
\sum_{i=-\infty}^{m-1}\frac{1}{2}\left(\Sigma^{>,<}(t_{n},t_{i})G^{A}(t_{i},t_{m})+\Sigma^{>,<}(t_{n},t_{i+1})G^{A}(t_{i+1},t_{m})\right)\right]\nonumber \\
F_{2}^{>,<}(t_{n},t_{m}) & =  +i\Delta t\left[\sum_{i=-\infty}^{n-1}\frac{1}{2}\left(G^{R}(t_{n},t_{i})\Sigma^{>,<}(t_{i},t_{m})+G^{R}(t_{n},t_{i+1})\Sigma^{>,<}(t_{i+1},t_{m})\right)\right.
\notag\\
&
\left.
+
\sum_{i=-\infty}^{m-1}\frac{1}{2}\left(G^{>,<}(t_{n},t_{i})\Sigma^{A}(t_{i},t_{m})+G^{>,<}(t_{n},t_{i+1})\Sigma^{A}(t_{i+1},t_{m})\right)\right],\mylabel{eq:KB_integral_trap_form}
\end{align}
and use these new expressions at every place where the evaluation of $F_{1,2}^{>,<}$ were performed in the previous discussion.
\section{Equilibrium Green's functions Calculations}
\mylabel{sec:sup:spectral}\mylabel{sec:eq_Gfs}
To obtain the equilibrium Green's functions we follow a similar approach to \cite{BanerjeeAltman2016}, and incorporate the conditions of equilibrium into the saddle point equations \eqn{eqn:Sg_Sl_def}. First, we assume time translation symmetry, which makes every two-point  correlator a function of $t=t_{1}-t_{2}$. Therefore, we can write \eqn{eqn:Sg_Sl_def} as
\begin{eqnarray}\mylabel{eqn:Sigma_great_less_SYKq_t}
\Sigma^{>}_s(t) & = & J_s^{2}G^{>}_s(t)^{q_s}G^{<}_s(-t)^{q_s-1}+(\sqrt{p})^{s}V^2G_{\bar{s}}^>(t_1,t_2)\nonumber \\
\Sigma^{<}_s(t) & = & J_s^{2}G^{<}_s(t)^{q_s}G^{>}_s(-t)^{q_s-1}+(\sqrt{p})^{s}V^2G_{\bar{s}}^<(t_1,t_2),
\end{eqnarray}
where we have made the $V$-term time independent, and introduced the symbols -- $J_s= J^2\ (t_\psi^2)$, $q_s= 2\ (1)$, $\bar{s}=\psi\ (c)$ when $s=c\ (\psi)$. The exponent of the factor $\sqrt{p}$ takes the value $+1$ when $s=c$ and $-1$ when $s=\psi$. To calculate the Green's functions for the (dis)connected system we set $V=(1)\ 0$. Using the relations between $G^>$, $G^<$ and $G^K,\ G^R,\ G^A$ (see \seccite{sec:eq_Gfs}) and the condition \eqn{eqn:FDT}, we have
\begin{eqnarray}\mylabel{eqn:Gg_Gl_eq_simp}
G^{>}(\omega) & = & (G^{K}+(G^{R}-G^{A}))/2=(\ci2\tanh(\beta\omega/2)+2\ci)\Im[G^{R}(\omega)]/2=-\ci2\pi n_{F}(-\beta\omega)\rho(\omega)\notag\\
G^{<}(\omega) & = & (G^{K}-(G^{R}-G^{A}))/2=(\ci2\tanh(\beta\omega/2)-2\ci)\Im[G^{R}(\omega)]/2=\ci2\pi n_{F}(\beta\omega)\rho(\omega),
\end{eqnarray}
where $\rho_s(\omega)$ are the spectral functions for the respective fermion flavors. Writing $G^{>,<}(t)$ in terms of $G^{>,<}(\omega)$ in \eqn{eqn:Sigma_great_less_SYKq_t} and using the definition for retarded function $\Sigma^{R}(t)=\Theta(t)[\Sigma^{>}(t)-\Sigma^{<}(t)]$, we find $\Sigma^{R}(\omega)$ to be
\bea\mylabel{eqn:sup:Sigma_wplus_GBA}
\Sigma_{s}(\omega^+)
=-\ci\int\limits_0^{\infty}\D{t}\ 
e^{\ci\omega t}
\left[
J_s^2
\left\{
 n_{1s}^{q_s-1}(t)
 n_{2s}^{q_s}(t)
+
n_{3s}^{q_s-1}(t)
n_{4s}^{q_s}(t)
\right\}
+
(\sqrt{f})^sV^2
\left\{
 n_{1s}^{r-1}(t)
 n_{2\bar{s}}^r(t)
+
n_{3s}^{r-1}(t)
n_{4\bar{s}}^r(t)
\right\}
\right],
\eea
where $n_{(1-4)s}(t)$ are defined as
\bea\mylabel{eqn:sup:nint_def}
\begin{array}{ll}
n_{1s}(t)=\int\limits_{-\infty}^{+\infty}\D{\Omega_k}\ \rho_s(\Omega_k)n_F(-\Omega_k) e^{+\ci\Omega_k t},&
n_{2s}(t)=\int\limits_{-\infty}^{+\infty}\D{\Omega_k}\ \rho_s(\Omega_k)n_F(\Omega_k) e^{-\ci\Omega_k t}\\
n_{3s}(t)=\int\limits_{-\infty}^{+\infty}\D{\Omega_k}\ \rho_s(\Omega_k)n_F(\Omega_k) e^{+\ci\Omega_k t},&
n_{4s}(t)=\int\limits_{-\infty}^{+\infty}\D{\Omega_k}\ \rho_s(\Omega_k)n_F(-\Omega_k) e^{-\ci\Omega_k t}.
\end{array}
\eea
The integro-differential equations in \eqn{eq:KB_eq_1_sec}, under the assumptions of equilibrium, reduces to the Dyson equations for both the fermion flavors and takes the form
\bea\mylabel{eqn:sup:Dyson_w}
G_s^R(\omega)=[\omega+\mu-\Sigma_s(\omega^+)]^{-1},
\eea
where $G_s^R$ is the retarded Green's function. The spectral function appearing in \eqn{eqn:Gg_Gl_eq_simp} and \eqn{eqn:sup:Sigma_wplus_GBA} is related to the retarded Green's function as
\bea\mylabel{eqn:sup:spectral_func_def}
\rho_s(\omega)=-\frac{1}{\pi}\Im \left[G_s^R(\omega)\right].
\eea
Using \eqn{eqn:sup:Sigma_wplus_GBA}, \eqn{eqn:sup:Dyson_w} and \eqn{eqn:sup:spectral_func_def} we iteratively solve for the spectral function $\rho_s(\omega)$. The iterative process is terminated when we have converged to a solution for $\rho_s(\omega)$ with sufficient accuracy. Using the converged value of $\rho_s(\omega)$ the equilibrium versions of $G^>(t)$ and $G^<(t)$ can be obtained by using the left most equations in \eqn{eqn:Gg_Gl_eq_simp} and then taking a Fourier transform.
\section{Time dependent expectation values}
\def\schrodinger{Schr$\ddot{\textup{o}}$dinger}
\mylabel{sec:neq_observable}
The time dependent expectation values for the energy components $E_c$,$E_\psi$, $E_{c\psi}$ etc. discussed  in the main text are derived using the identity\mycite{KamenevBook}
\bea\mylabel{eqn:neq_Ot_app}
\langle\hat{O}(t)\rangle=\lim_{\eta\to 0}\frac{\ci}{2}\frac{\delta Z_{neq}[\eta]}{\delta \eta(t) },
\eea
where $Z_{neq}$ is the generating functional, introduced in \seccite{sec:neq_action}, now containing an additional source field $\eta(z)\hat{O}(t)$ with $\eta(z)$ defined as 
\begin{align}\mylabel{eqn:Vz_app}
\eta(z)=\begin{cases}
+\eta(t)&\textup{when } z\in +ve\ \textup{branch}\\
-\eta(t)&\textup{when } z\in -ve\ \textup{branch}
\end{cases},
\end{align}
and $\hat{O}$ is an operator whose expectation value we want to evaluate. The following source fields are used to derive the time dependent energy components
\begin{enumerate}
\item $(1+\eta(z)){1\over(2\Lc)^{3/2}}\sum\limits_{\mathclap{ijkl}} 
J_{ijkl}\cb_{i}\cb_{j} c_{k}c_{l}$, $(1+\eta(z)){1\over \LPsi^{1/2}}\sum\limits_{\mathclap{\alpha\beta}} 
t_{\alpha\beta}^\psi\Psib_{\alpha}\psi_{\beta}$ for evaluating $\langle\cH_{c}(t)\rangle$ $\equiv E_{c}(t)$ and $\langle\cH_{\psi}(t)\rangle$ $\equiv E_{\psi}(t)$ respectively.
\item $f(z)(1+\eta(z))
\sum\limits_{\mathclap{i\alpha}}
 V_{i\alpha}\cb_{i}\psi_{\alpha}+ \mbox{h.~c.}$, for evaluating $\langle \cH_{c\psi}(t)\rangle$ $\equiv E_{c\psi}(t)$.
\end{enumerate}
Using the identity in \eqn{eqn:neq_Ot_app} we calculate the disorder averaged time dependent expectation values for the individual parts of the Hamiltonian defined in \eqn{eq.Model}, and find
\begin{align}
E_c(t)=\langle\cH_{c}(t)\rangle=&(-\ci)\frac{\Lc J^{2}}{2}\int_{-\infty}^{t}\dt_{1}\left[\Gc^{>}(t,t_{1})^{2}\Gc^{<}(t_{1},t)^{2}-\Gc^{<}(t,t_{1})^{2}\Gc^{>}(t_{1},t)^{2}\right]\nonumber\\
E_\psi(t)=\langle\cH_{\psi}(t)\rangle=&(-\ci){\LPsi t_\psi^{2}}\int_{-\infty}^{t}\dt_{1}
\left[\GPsi^{>}(t,t_{1})\GPsi^{<}(t_{1},t)-\GPsi^{<}(t,t_{1})\GPsi^{>}(t_{1},t)\right]
\nonumber\\
E_{c\psi}(t)=\langle \cH_{c\psi}(t)\rangle=&(-\ci){\sqrt{\Lc\LPsi}V^2}
\int_{-\infty}^{t}\dt_{1}f(t)f(t_{1})\left[G_{\psi}^{>}(t,t_{1})G_{c}^{<}(t_{1},t)+G_{c}^{>}(t,t_{1})G_{\psi}^{<}(t_{1},t)\right.\nonumber\\
&\left.-G_{\psi}^{<}(t,t_{1})G_{c}^{>}(t_{1},t)-G_{c}^{<}(t,t_{1})G_{\psi}^{>}(t_{1},t)\right].
\mylabel{eq:Energy_neq_t_sec}
\end{align}
Using the above quantities, an excitation energy produced during the quench can be defined as
\begin{align}
\Delta E(\tau)=E(\tau)-\langle \mc{H}(\tau)\rangle_{T_i}\mylabel{eqn:deltaEdef},
\end{align}
where $\langle \mc{H}(\tau)\rangle_{T_i}$ is the thermal expectation of energy for the final Hamiltonian $\mc{H}(\tau)$ at the initial temperature $T_i$ from which the quench began.

An additional observable can be defined corresponding to the occupation density $n_\psi(\epsilon,t)$ of a $\psi$-fermion having energy $\epsilon$. This is possible since, the $\psi$ fermions in the non-interacting model are connected via random hoppings, (see \eqn{eq.SYK_SI}) and hence can be diagonalized to yield a set of eigenstate fermions which we label by $\epsilon$. The Kadanoff-Baym equations satisfied by the Green's functions for the $\psi$ fermions are 
\begin{eqnarray}
(\ci\partial_{t_{1}}-\epsilon+\mu)G^{>,<}_{\psi}(\epsilon;t_{1},t_{2}) & = & \int_{-\infty}^{t_{1}}\Sigma^{R}_{\psi}(\epsilon;t_{1},t)G^{>,<}_{\psi}(\epsilon;t,t_{2})\dt+\int_{-\infty}^{t_{2}}\Sigma^{>,<}_{\psi}(\epsilon;t_{1},t)G^{A}_{\psi}(\epsilon;t,t_{2})\dt\nonumber \\
(-\ci\partial_{t_{2}}-\epsilon+\mu)G_{\psi}^{>,<}(\epsilon;t_{1},t_{2}) & = & \int_{-\infty}^{t_{1}}G_{\psi}^{R}(\epsilon;t_{1},t)\Sigma_{\psi}^{>,<}(\epsilon;t,t_{2})\dt+\int_{-\infty}^{t_{2}}G_{\psi}^{>,<}(\epsilon;t_{1},t)\Sigma_{\psi}^{A}(\epsilon;t,t_{2})\dt\nonumber\\\label{eq:KB_phi_eq_1},
\end{eqnarray}
where
\begin{eqnarray}
\Sigma_\psi^{R}(\epsilon;t_{1},t_{2}) & = & \Theta(t_{1}-t_{2})[\Sigma_\psi^{>}(t_{1},t_{2})-\Sigma_\psi^{<}(t_{1},t_{2})]\nonumber \\
\Sigma_\psi^{A}(\epsilon;t_{1},t_{2}) & = & \Theta(t_{2}-t_{1})[\Sigma_\psi^{<}(t_{1},t_{2})-\Sigma_\psi^{>}(t_{1},t_{2})].
\end{eqnarray}
The number density for the $\psi$ fermions can be obtained from the Keldysh Green's function using 
\bea\mylabel{eqn:nphi_t}
n_{\psi}(\epsilon,t)=(1-\ci G^K_{\psi}(\epsilon;t,t))/2,
\eea
where $G^K_{\psi}(\epsilon;t,t)=G^>_{\psi}(\epsilon;t,t)+G^<_{\psi}(\epsilon;t,t)$.

\section{Sudden Quench}

We now briefly discuss some of the results obtained from the sudden quench of the interacting as well as the non-interacting models, and also provide a comparison between the two cases.
\subsection{Energy time-series}\mylabel{subsec:E_consv}
\begin{figure}[t]
\includegraphics[scale=1]{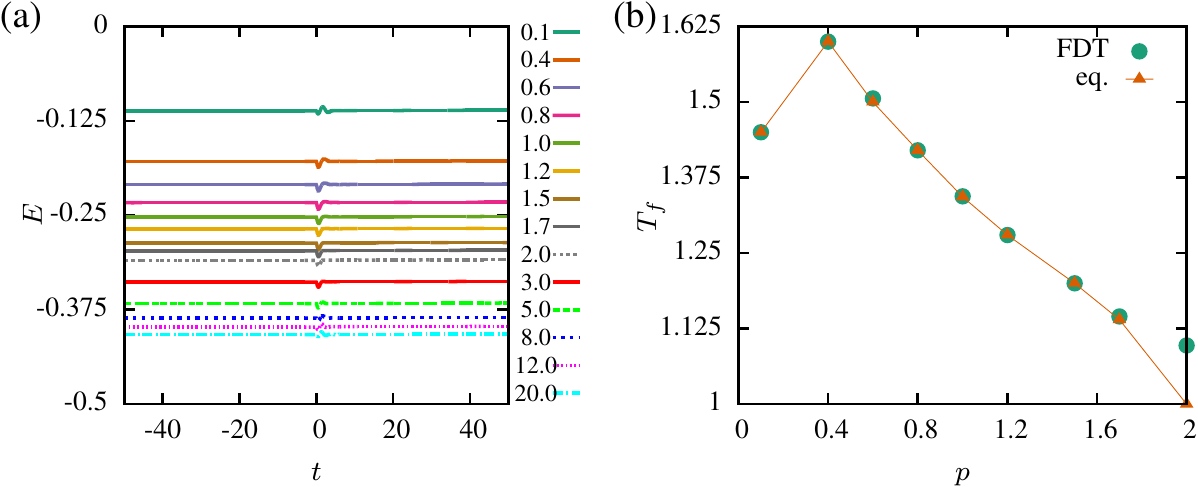} 
\caption{{\bf Energetics for the interacting model quench:} (a) Total energy versus time, showing the sudden quench process to be iso-energetic. The small fluctuations near $t=0$ are an artifact of numerics and reduce when the resolution is increased. (b) Final temperature obtained using the fluctuation dissipation theorem (FDT) shown as function of site-fraction $p$. The FDT temperature is in agreement with the values obtained from equilibrium calculations (marked with triangles) which were performed assuming an iso-energetic quench process.}
\mylabel{fig:Etot_time_series}\mylabel{fig:Tf_vs_p}
\end{figure}

We evaluate the total energy for the system, i.e.
\begin{align}
E(t)=E_c(t)+E_\psi(t)+E_{c\psi}(t),
\end{align}
as a function of time for various values of site-fraction $p$. The time-series data is shown in \fig{fig:Etot_time_series}(a). Interestingly, the sudden change of $f(t)$ in \eqn{eq.Model}, from 0 to 1, is an iso-energetic process and occurs without any work input to the system. The tiny fluctuations near $t=0$ are an artifact of discretization and reduces as the resolution of the time grid (see \fig{fig:KB_eqn_t1t2_palne_app}) is increased. We verify the iso-energetic nature of the quench by comparing the final temperature obtained from the FDT relation in \eqn{eqn:FDT}, with the temperature obtained from equilibrium calculations. The temperature in the latter case was found by solving for the value that produced the same energy for the connected $c-\psi$ system equal to the initial disconnected one. As shown in \fig{fig:Tf_vs_p}(b), the temperature vs. $p$ curves obtained for the above two cases are qualitatively similar proving that the sudden quench of the interacting model is indeed iso-energetic.\ttd{ The sudden quench of the non-interacting model is an iso-energetic process as well. However, unlike the interacting case, the non-interacting energy time-series is independent of $p$. This is consistent with the model, since in the initial disconnected system the $c$ and $\psi$ fermions are both described by a semi-circular density of states scaled appropriately by factors of $1/(1+p)$ and $p/(1+p)$ respectively, which cancel out when the energies of the two fermion flavors are added.}


\subsection{Thermalization of the occupation function and $E_{c\psi}$}
\mylabel{sec:int_sudden_results}

The thermalization behavior for the interacting model is shown in \fig{fig:thermalization_int}. The energy associated with the bonds between the $c$ and $\psi$ sites are shown as a function of time for various values of site-fraction $p$. We find for small values of $p$, i.e. $p<1$, the energy reaches the equilibrium value (shown with arrow heads, see \fig{fig:thermalization_int}(a)) very rapidly. This is expected, since the $p<1$ phase after the quench is a SYK-type NFL, which is known to be a fast thermalizer. On the other hand, for values of \ttd{$p>3.0$} the approach to the equilibrium state slows down rapidly, as demonstrated by the negative slope of $E_{c\psi}-t$ curves, for $t>0$, in \fig{fig:thermalization_int}(b). Again, this behavior is consistent with the equilibrium model since at larger values of $p$ a FL-state is expected to dominate the thermalization time scales of the system. The same phenomenon is observed in the distribution functions of the $\psi$ fermions as well and which is reported in the main text. For small values of $p$, like $p=0.1$, (see \fig{fig:SuddenQuench}(e) top) the distribution function, evaluated sometime after the quench, matches well with that obtained from a equilibrium calculation using the temperature determined via energy conservation. However, at a large value of $p=8.0$, see \fig{fig:SuddenQuench}(e) bottom, the match is poor due to the slowed down approach towards equilibrium.

Moving on to the sudden quench of the non-interacting model, we find a drastically different behavior, in that the system completely fails to thermalize. \Fig{fig:thermalization_nI}(a) demonstrates this feature very clearly, the energy $E_{c\psi}$ reaches a steady state value, right after the quench, which is very different from the expected equilibrium value. Unlike the interacting model for large $p$ values, where approach to equilibrium was simply slowed down, in this situation we have a complete halt on equilibration. The above fact is further supported by the form of the $\psi$-fermion distribution function as shown in \fig{fig:thermalization_nI}(b) and (c). Irrespective of the value of site-fraction $p$, the distribution function fails to converge towards its equilibrium counterpart. This is in sharp contrast to the interacting model, where we had a regime of $p$ values for which the distribution function converged really well with the equilibrium results.

In summary, we find that the interacting model can be distinguished from the non-interacting model by studying their thermalization behavior. In the case of interacting model, the system either thermalizes rapidly or slowly depending on the regime of $p$ values we are looking at. On the other hand, the non-interacting model ceases to thermalize at all and reaches a steady state, far away from equilibrium, irrespective of the value of site fraction $p$.

\subsubsection{Exact diagonalization study of thermalization of $E_{c\psi}$ at finite $N$}
\mylabel{sec:ED_sudden_results}
As discussed in the main text, one can ask whether the crucial features of the large-$N$ non-equilibrium dynamics, e.g. the fast and slow thermalization in the NFL and FL, respectively, persist at finite $N$. To answer this, we perform numerical exact diagonalization (ED) studies of the sudden quench in our model. We take the $T=0$ direct product ground state $|\Psi_0\rangle=|NFL\rangle \otimes |FL\rangle$ as the initial state of the initially decoupled system, where $|NFL\rangle$ is the ground state of the SYK model (\eqn{eq.SYK}) with $N$ sites and $|FL\rangle$ the ground state of the lead fermions described by \eqn{eq.Anderson} for a particular disorder realization. As in the large-$N$ calculations, we switch on the coupling term $\mathcal{H}_ {c\psi}$ (\eqn{eq.Coupling}) at $t=0$. After the quench, $|\Psi_0\rangle$ is no longer an eigenstate of the Hamiltonian $\mathcal{H}$ (\eqn{eq.Model}) and its time evolution is given by $|\Psi(t)\rangle=e^{-\ci\mathcal{H}t}|\Psi_0\rangle $. The disorder-averaged $E_{c\psi}(t)=\overline{\langle \Psi_0(t)| \mathcal{H}_{c\psi}|\Psi_0(t)\rangle }$ is shown in \fig{fig:thermalization_int}(c) and in \fig{fig:SuddenQuench}(h) (main text). The disorder average is taken over $300$ disorder realizations. To obtain the time-evolution, the initial state	$|\Psi_0\rangle=\sum_n c_n|\phi_n\rangle$ is written in terms of eigenstate $|\phi_n\rangle$ of the post quench Hamiltonian $\mathcal{H}$, so that 	$|\Psi(t)\rangle=\sum_n c_n e^{-\ci E_n t}|\phi_n\rangle$, where $c_n = \langle \phi_n | \Psi_0 \rangle $ and $E_n$'s are eigen energies of $\mc{H}$. 

The ED results are obtained at half filling for total system size $N+M=16$. The plots (\fig{fig:thermalization_int}(c) and in \fig{fig:SuddenQuench}(h)) are shown for $p=0.33$ (i.e. $N=12, M=4$) and for $p=3.0$ (i.e. $N=4, M=12$). According to eigenstate thermalization hypothesis (ETH) \cite{DAlessio2016}, if the coupled system thermalizes, then, in the long-time limit, $E_{c\psi}$ is expected to reach the value, $E^d_{c\psi}=\overline{\Tr(\rho_{d} H_{c\psi})}$, given by the diagonal ensemble corresponding to the density matrix $\rho_{d}=\sum_n |c_n|^2 |\phi_n\rangle \langle \phi_n | $. The expectation value in the diagonal ensemble matches with the thermal expectation, that is obtained from microcanonical ensemble, within a sub-extensive correction. We have checked that there is indeed a sub-extensive difference ($\sim 1\%$ ) between the values of $E_{c\psi}$ in the diagonal and microcanonical ensembles . Hence, to avoid finite size effects we compare long time $E_{c\psi}(t)$ with the diagonal expectation value. As shown in 
\fig{fig:thermalization_int}(c), for $p=0.33$, $E_{c\psi}(t)$ reaches diagonal ensemble expectation value much faster than that for $p=3.0$. Unlike the large-$N$ results in Fig.\ref{fig:SuddenQuench}(e), we could not study the thermalization deeper in the FL phase due to the limitation of system sizes ($N$ and $M$) in ED. Nevertheless, the difference of thermalization rates in the NFL and FL is quite evident even from the comparison of $E_{c\psi}(t)$ for $p=0.33$ and $p=3$.
 
\begin{figure}
\includegraphics[scale=1]{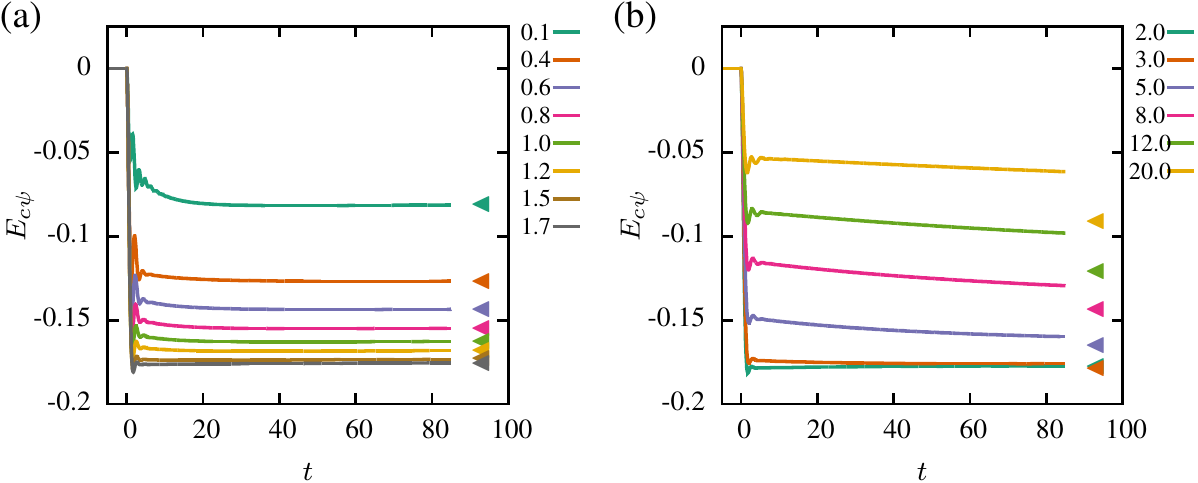}
\includegraphics[scale=0.77]{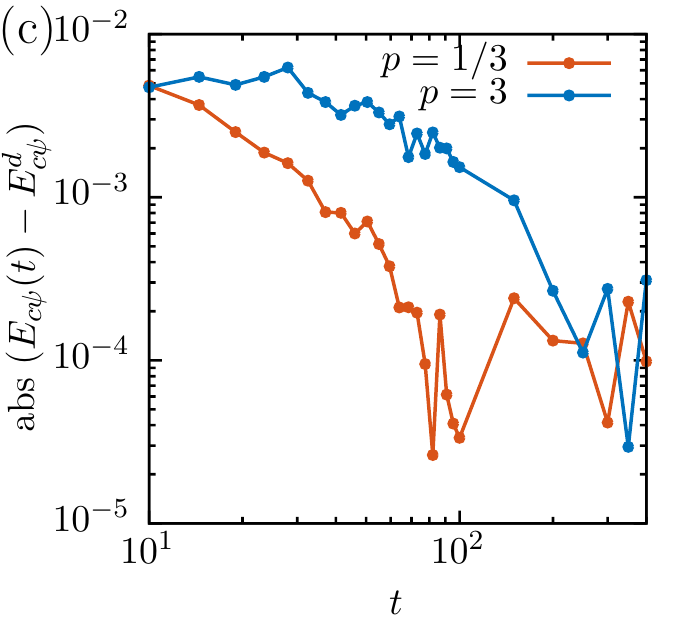} 
\caption{{\bf Thermalization behavior of the interacting model:} The energy $E_{c\psi}$, associated with the bonds between $c$ and $\psi$ fermions, shown as a function of time for site-fractions $p=0.1-1.7$ in (a) and $p=2.0-20.0$ in (b). When $p\lesssim 3.0$ the energy rapidly approaches the expected equilibrium value (arrow heads) after the quench and the system gets equilibrated very quickly. On the other hand, when $p>3.0$, the rate of equilibration gets slowed down drastically and the energy takes a much longer time to reach its equilibrium value. (c) The absolute difference between the bond energy $E_{c\psi}(t)$ and the diagonal ensemble expectation value of the bond energy  $E^d_{c\psi}$, obtained via exact diagonalization (ED) for total system size $N+M=16$, is shown as a function of time($t$) for site fraction $p=1/3$ and $p=3$. Clearly, the difference $|E_{c\psi}(t)-E^d_{c\psi}|$ for $p=1/3$ approaches \emph{zero} faster than the $p=3$ case, indicating that thermalisation happens faster for $p=1/3$ than $p=3$.}\label{fig:thermalization_int}
\end{figure}

\begin{figure}
\includegraphics[scale=1]{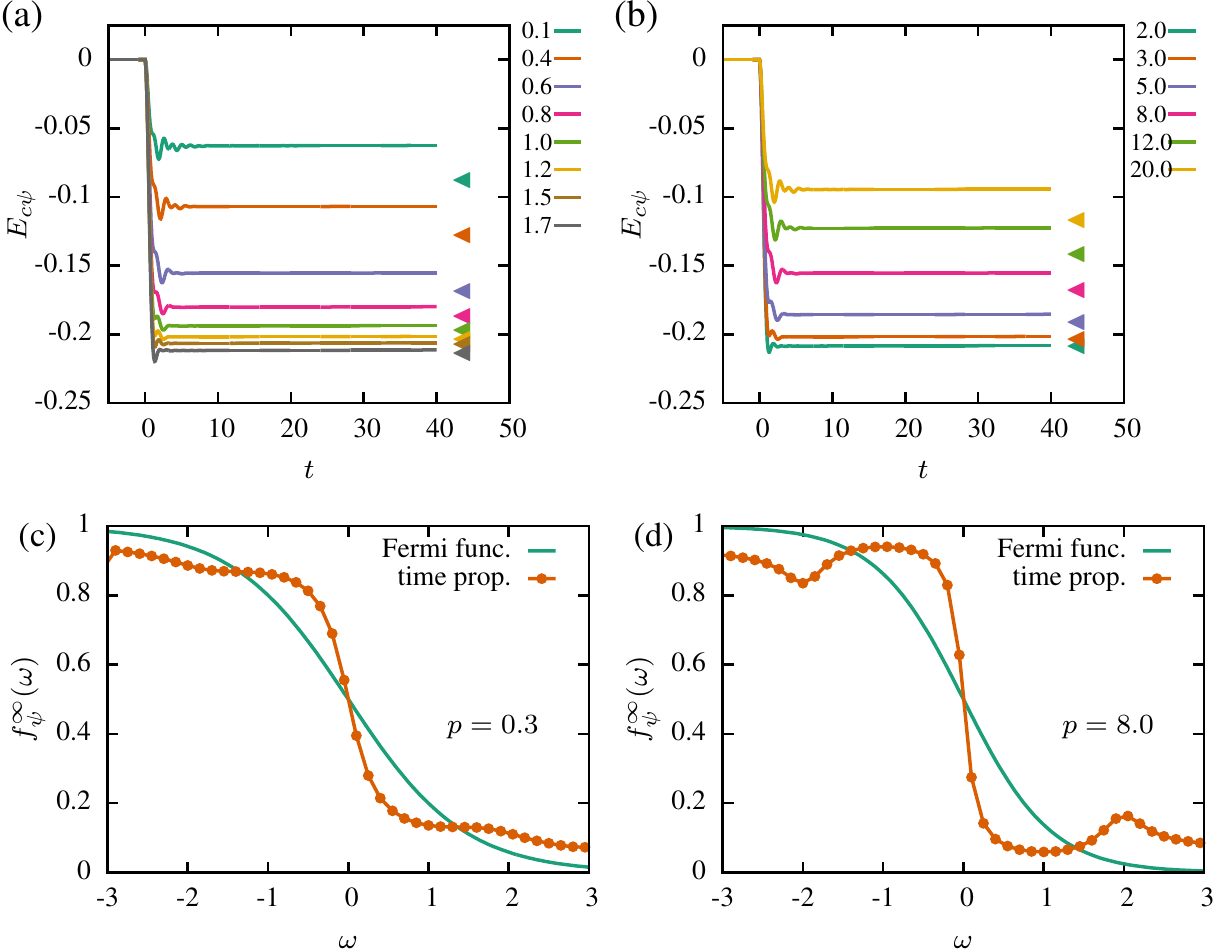}
\caption{{\bf Absence of thermalization in the non-interacting model:} (a)-(b) The energy density $E_{c\psi}$, associated with the $c-\psi$ bonds, plotted as a function of time. The energy reaches a steady-state value far from the equilibrium value (shown with a arrow head) right after the quench, and unlike the interacting model, this complete halt of thermalization occurs for all values of $p$. (c)-(d) The steady-state occupation function $f^\infty_\psi(\omega)$ for the $\psi$ fermions shown for site-fraction $p=0.3$ and $8.0$ respectively. The failure to thermalize to an equilibrium ensemble causes the occupation function (points) to deviate appreciably from the Fermi-function (solid line).}
\label{fig:thermalization_nI}
\end{figure}

\subsection{Collapse-revival to prethermal transition in the non-interacting model}
\mylabel{sec:nI_sudden_results}
\begin{figure}[!htb]
\ifdefined\showfigures
\centering
\includegraphics[scale=1]{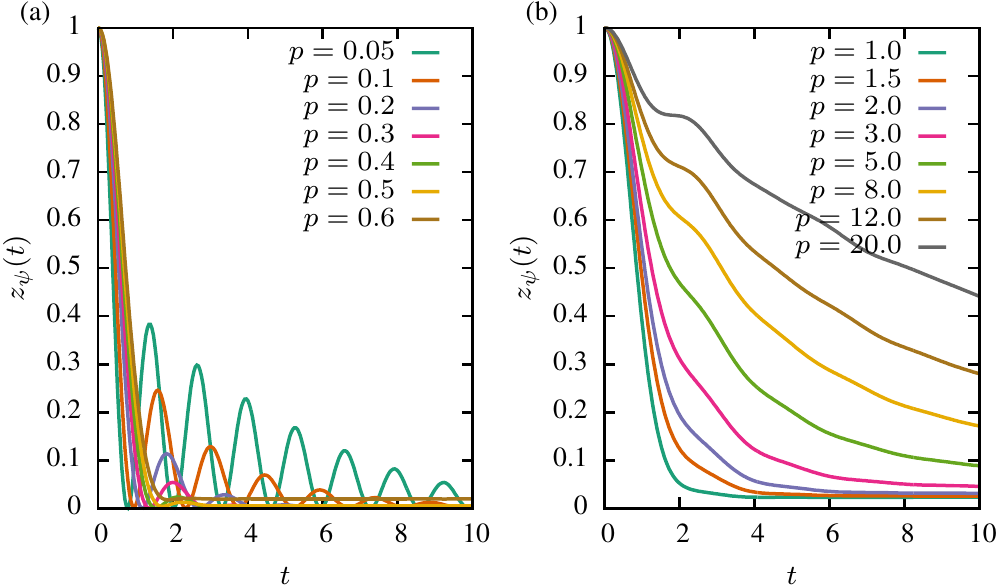} 
\includegraphics[scale=1]{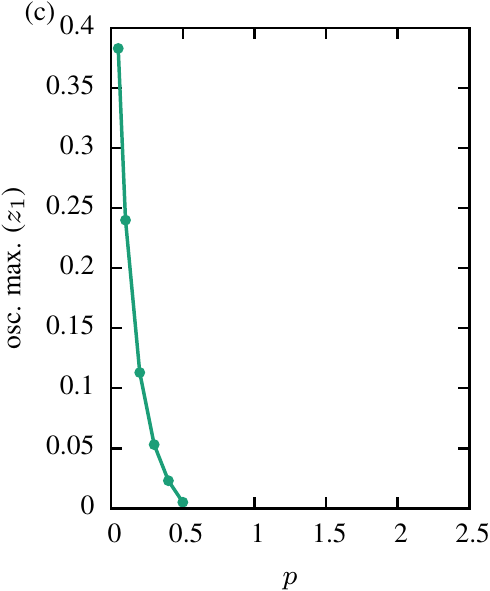} 
\caption{{\bf Results for the sudden quench of the non-interacting model:}  (a) The jump $z_\psi(t)$ in the fermion distribution shows oscillations similar to the interacting quench given in \fig{fig:SuddenQuench}(a). (b) Plot of $z_\psi(t)$ for higher values of $p$, showing the pre-thermal plateaus similar to \fig{fig:SuddenQuench}(b). (c) Plot of first maxima amplitude $z_1$, in fig.(a), as a function of $p$ showing a $p^{dyn}_c\sim 0.5$.}\label{fig:q11_sudden_quench}
\end{figure}
\fi
In this subsection we discuss the sudden quench of the non-interacting defined in \seccite{sec:nI_model}. The results of this exercise are given in \fig{fig:q11_sudden_quench}. Indeed, we find the same qualitative behavior as the interacting case, however with rescaled values for $p_c^{dyn}$. In fact, the way the oscillations disappear as a function of $p$ (see \fig{fig:q11_sudden_quench}(c)) are very similar to the interacting case (see inset in \fig{fig:SuddenQuench}(a)) with $p_c^{dyn}\approx 0.5$ in this case. The oscillations in $z_\psi$ that appear in \fig{fig:SuddenQuench}(a) are also reproduced (see \fig{fig:q11_sudden_quench}(a)) and occur when $p\leq {p_c^{dyn}}$. The prethermal plateaus and long thermalization times shown in \fig{fig:SuddenQuench}(b) are obtained at higher but different values of $p$, see \fig{fig:q11_sudden_quench}(b). Hence overall, the relaxation features of the fermion distribution function associated with the $\psi$ fermions, for the interacting model, are also observed in the sudden quench of the non-interacting model. Furthermore, the value of $p_c^{dyn}$ for this case can be explained by closing of a soft-gap in the spectral function of the fermions, see \seccite{sec:gap_closing} for details.

\section{Slow quench}
In this section, we discuss the details of the results that we obtained from the slow quench studies of both the non-interacting and the interacting models.
\subsection{Effect of ramp shapes on the interacting model quench}\mylabel{subsec:rampshapeseffect}
We join the SYK $c$-fermions with $\psi$-fermions over a  time-period  $\tau$ using ramps having various shapes and heights.
\begin{figure}[!htb]
\centering
\includegraphics[scale=1]{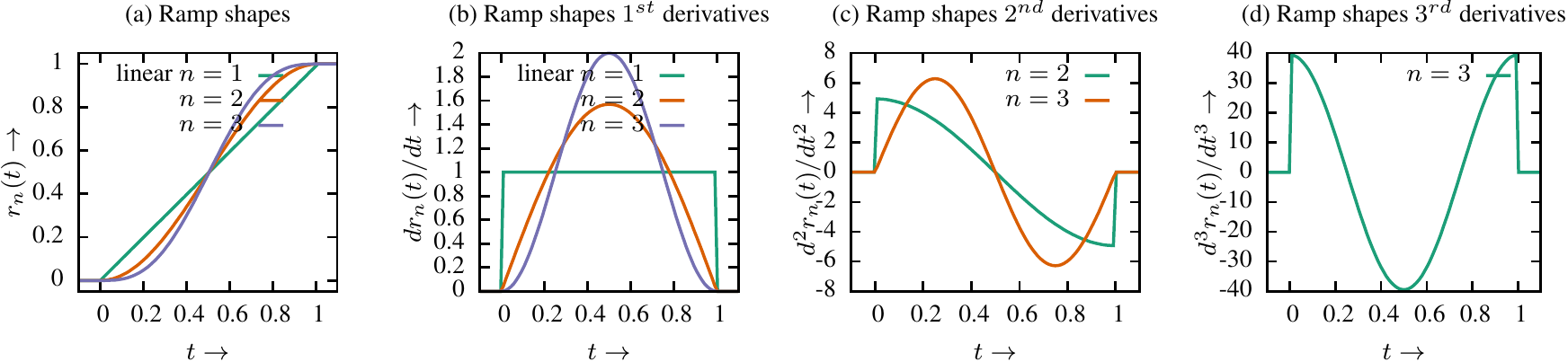} 
\caption{{\bf Ramp shapes and their derivatives:} (a) Ramp shapes $r_{n}(t)$ that were used to join the $c$ and $\psi$ fermions together. Figures (b)-(d) show the first, second and third derivatives of the ramp shapes respectively. The ramps are classified by the order of their first discontinuous derivative. The mathematical expressions for the ramp shapes are given in \eqn{eqn:ramps}.}\label{fig:ramp_shapes}
\end{figure}
The ramp shapes that we use, and shown in \fig{fig:ramp_shapes}, are chosen following ref.\onlinecite{Eckstein2010}, and are classified in the order of their smoothness. In particular, $r_n(t)$ will have its $n$-th derivative to be discontinuous at the start and end of the  ramp (\fig{fig:ramp_shapes}(b)-(d)). The ramp functions for $n=1,2,3$ are given below
\begin{subequations}
\begin{align}
r_1(x)&=x\\
r_2(x)&=\frac{(1-\cos(\pi x))}{2}\\
r_3(x)&=\frac{(\pi x-\cos(\pi x)\sin(\pi x))}{\pi}.
\end{align}\mylabel{eqn:ramps}
\end{subequations}
\begin{figure}[h!]
\ifdefined\showfigures
\centering
\includegraphics[scale=1]{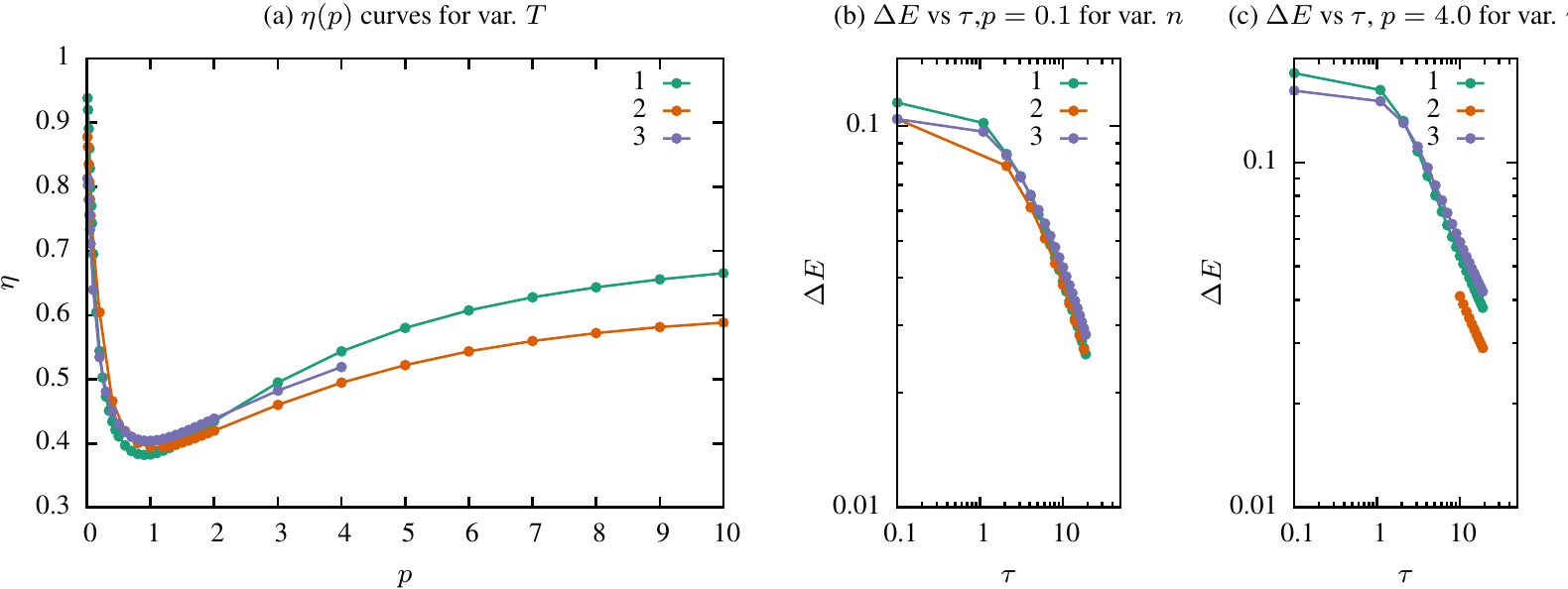}  
\fi
 \caption{{\bf Results for different ramp shapes:} (a) $\eta-p$ curves at $T=0.05$ for ramp shapes given in \fig{fig:ramp_shapes}(a) showing a weak dependence on ramp-shape around the critical $p$ value with the effect getting stronger away from it. (b) $\Delta E-\tau$ log-log scale plot for $p=0.1$ for various ramp shapes showing  a moderate dependence on $n$. (c) $\Delta E-\tau$ log-log scale plot for $p=4.0$ for various ramp shapes showing  a strong dependence on $n$.}
\label{fig:Vf_1_vs_T_0.05_vs_rs}
\end{figure}
The $\eta-p$ curves, calculated at an initial temperature $T_i=0.05$, for the ramp shapes in \eqn{eqn:ramps} are shown in \fig{fig:Vf_1_vs_T_0.05_vs_rs}(a). We find that there exists a dependence on ramp-shapes. However,  this dependence diminishes drastically around the transition point $p=1$, indicating that the intrinsic properties of the critical point are becoming more prevalent. Deep within the phases, the dependence on ramp-shape is the strongest, with the the stronger effect manifesting in the FL limit($p\to\infty$) (see \fig{fig:Vf_1_vs_T_0.05_vs_rs}(b),(c)). 

\subsubsection{Effect of temperature on the interacting model quench}
\begin{figure}[!htb]
\ifdefined\showfigures
\centering
\includegraphics[scale=0.7]{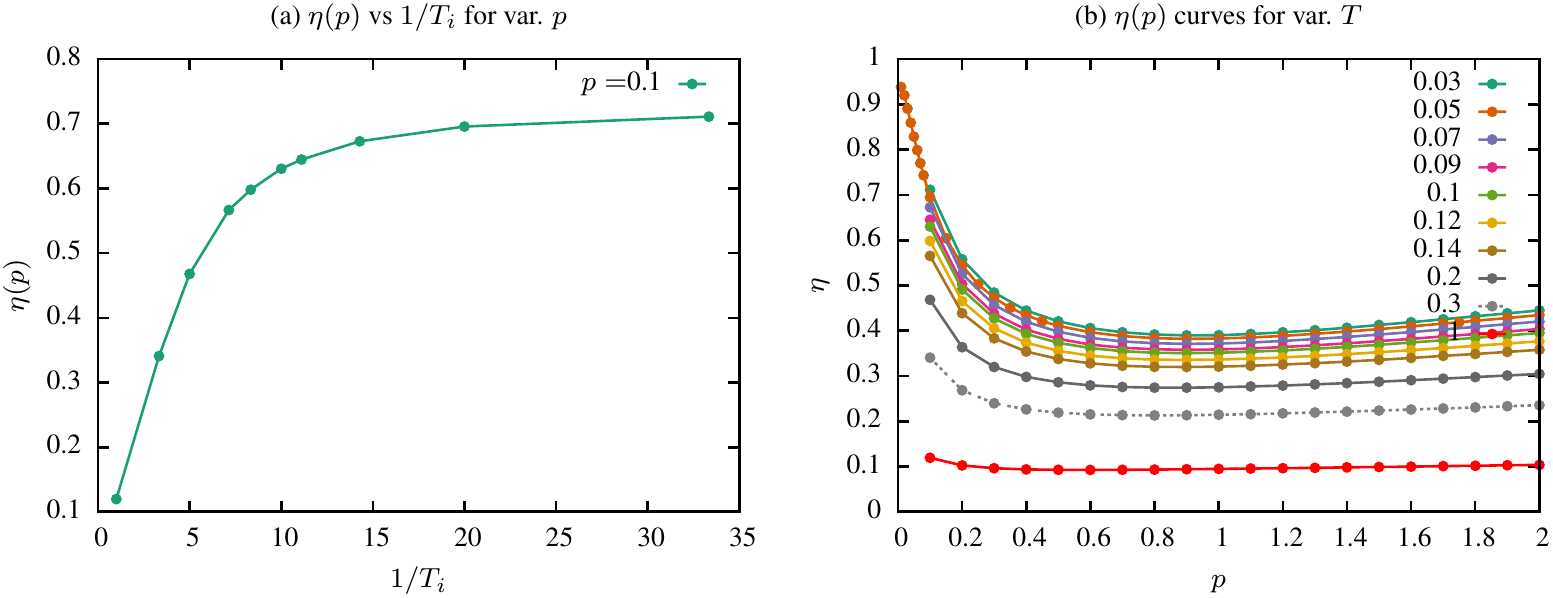}
\fi
\caption{{\bf Temperature dependence of powerlaw exponent:} (a) The exponent $\eta(p=0.1)$ as a function of inverse initial temperature $1/T_i$ showing a convergence to a value of $0.7$ at low temperatures. (b) $\eta-p$ curves for various $T_i$, increasing temperature gradually washes off the minimum at $p_{crit}$ as well as the dependence on $p$. At lower temperatures, the curves start to converge to a single function, in accordance with (a).}
\label{fig:linear_ramp_results_Vf_1_vs_T}
\end{figure}
In the case of the interacting model the initial temperature $T_i=0$ is not readily accessible, therefore we ask how does the exponent $\eta$ change as a function $T_i$ for any given $p$ value? The behavior of $\eta(p,T_i)$ as a function of $1/T_i$, for $p=0.1$, is shown in \fig{fig:linear_ramp_results_Vf_1_vs_T}(a), and suggests that $\eta(p,T_i)$ converges to a finite value at low temperatures. Further, we obtain the $\eta-p$ curves (see \fig{fig:linear_ramp_results_Vf_1_vs_T}(b)) at various temperatures, and find that they indeed converge to a finite value as $T_i\to 0$. This suggests that the powerlaw behavior that we observe is a genuine property of the quantum ground states involved in the quench.

%

\fi
\end{document}